\documentclass[pdflatex,sn-mathphys-num]{sn-jnl}

\usepackage{graphicx}%
\usepackage{multirow}%
\usepackage{amsmath,amssymb,amsfonts}%
\usepackage{amsthm}%
\usepackage{mathrsfs}%
\usepackage[title]{appendix}%
\usepackage{xcolor}%
\usepackage{textcomp}%
\usepackage{manyfoot}%
\usepackage{booktabs}%
\usepackage{algorithm}%
\usepackage{algorithmicx}%
\usepackage{algpseudocode}%
\usepackage{listings}%

\theoremstyle{thmstyleone}%
\theoremstyle{thmstyletwo}%

\theoremstyle{thmstylethree}%

\begin{document}

\title[LLMs' Impacts on Software Development]{LLMs' Reshaping of People, Processes, Products, and Society in Software Development: A Comprehensive Exploration with Early Adopters}
\author*{\fnm{Benyamin} \sur{Tabarsi}}\email{btaghiz@ncsu.edu}
\equalcont{These authors contributed equally to this work.}
\author{\fnm{Heidi} \sur{Reichert}}
\equalcont{These authors contributed equally to this work.}
\author{\fnm{Sam} \sur{Gilson}}
\author{\fnm{Ally} \sur{Limke}}
\author{\fnm{Sandeep} \sur{Kuttal}}
\author{\fnm{Tiffany} \sur{Barnes}}

\affil{\orgname{North Carolina State University}, \orgaddress{\city{Raleigh}, \state{North Carolina}, \country{USA}}}


\abstract{
Large language models (LLMs) are rapidly reshaping software development, but their impact across the full software development lifecycle is underexplored. Existing work tends to focus on isolated activities such as code generation or testing, leaving open questions about how LLMs affect developers, processes, products, and the broader software ecosystem.

We address this gap through semi-structured interviews with sixteen early-adopter software professionals who integrated LLM-based tools into their day-to-day work in early to mid-2023. We treat these interviews as early empirical evidence and, in the discussion, compare participants’ accounts with recent work on LLMs in software engineering, noting which early patterns persist or shift. Using thematic analysis, we organize our findings around four dimensions: people, process, product, and society. Developers reported substantial productivity gains from reducing mundane tasks, streamlining search, and accelerating debugging, but also described a productivity–quality paradox: they frequently discarded generated code and shifted effort from writing code to critically evaluating and integrating it. LLM use was highly phase-dependent, with strong uptake in implementation and debugging but limited influence on requirements gathering and collaborative work.

Participants developed new competencies to use LLMs effectively, including prompt engineering strategies, multi-layered verification, and security-conscious integration to protect proprietary data. They also anticipated changes in hiring expectations, team practices, and computing education, while emphasizing that human judgment and foundational software engineering skills remain essential. Our findings, later echoed in large-scale studies, offer actionable implications for developers, organizations, educators, and tool designers seeking to integrate LLMs responsibly into professional software practice.}

\maketitle

\section{Introduction} 
Large language models (LLMs) are generative systems trained on extensive text and code corpora that can produce human-like responses to natural language prompts \cite{openai2023gpt, georgetown2023AI}. Publicly-available LLMs such as OpenAI ChatGPT\footnote{https://openai.com/chatgpt}, Anthropic Claude\footnote{https://claude.ai/}, and Google Gemini\footnote{https://gemini.google.com/} have rapidly reshaped workflows across multiple domains \cite {Hoff_Mok_Zinkula_2023, Mok_Zinkula_2023, Verma_2023}. 
Within software development, LLMs trained on vast code repositories, such as OpenAI's Codex \cite{ghcopilot_codex_1, ghcopilot_codex_2}, enable developers to generate, refactor, and debug code through natural language prompts. These tools streamline formerly manual tasks and have become central to developer productivity, collaboration, and innovation in software engineering \cite{chen2024opportunities, rane2023contribution}.

The rapid adoption of LLMs in software development has prompted ongoing debate about how they should be integrated into professional practice. Stack Overflow’s annual developer surveys report growth from 44\% to 76\% of developers using or planning to use AI tools between 2023 and 2024 \cite{stack_overflow_survey_2023, stack_overflow_survey_2024}, and enterprise deployments report substantial productivity gains \cite{github_accenture_2024}. However, concerns about code quality, security vulnerabilities, and impacts on developer learning have emerged alongside these benefits \cite{pearce2025asleep, sajadi2025llms}. 

Empirical work on LLMs in software engineering is still uneven. Existing studies predominantly address isolated activities, such as code generation \cite{chen2021evaluating}, testing \cite{yang2024evaluation}, or requirements engineering \cite{hemmat2025research}. Other work has examined LLM usage in software engineering through literature reviews \cite{hou2023large}, case studies \cite{belzner2023large}, empirical studies of forums \cite{chen2024empirical}, and comparison of general LLM tools and LLM-based agents \cite{jin2024llms}. Formal qualitative interviews with professional developers remain limited, and those that exist typically address narrow aspects such as security \cite{klemmer2024using}, trustworthiness \cite{rabani2023developers}, and user study evaluations of new tools \cite{pinto2024developer}, rather than tracing how LLMs impact people, processes, products, and organizational and societal issues across the software development lifecycle. Moreover, much of this empirical evidence was collected in the first wave of LLM adoption; assessing its continued relevance now requires explicitly relating early accounts to more recent studies.

To address this gap, we conducted interviews with 16 early-adopter software industry professionals who took the initiative to become educated about LLMs at their inception and began actively incorporating LLM-based AI tools into their daily workflows between November 2022 and April 2023. These early adopters, interviewed in mid-2023, provide critical insights into emerging patterns that subsequent large-scale studies also report \cite{cui2025effects, github_accenture_2024, stack_overflow_survey_2024}, which makes them particularly valuable for understanding how LLM integration has evolved in professional software development.
A further objective of this study is to compare these early-adopter accounts with the rapidly expanding literature on LLMs in software engineering. We revisit this comparison in the discussion to illustrate how early patterns have persisted, evolved, or been challenged by subsequent work.
Our analysis is organized around four dimensions that cover the software development lifecycle and its broader context: \textbf{people}, examining how LLMs influence individual developers and teams; \textbf{process}, analyzing changes in software engineering workflows; \textbf{product}, evaluating how LLMs contribute to software quality and innovation; and \textbf{society}, exploring the broader socioeconomic and ethical implications. This framework allows us to describe how LLMs have been reshaping software engineering practice, from individual task execution to organizational transformation.

\begin{itemize}

\item \textbf{People -- RQ1: How do LLMs affect developers?}

We examine how LLMs change day-to-day development work, focusing on perceived benefits, risks, and shifts in skills. Our study's participants reported clear gains in productivity by offloading mundane tasks, streamlining search, and utilizing boilerplate code. They also described LLMs as facilitating learning through personalizing education and explaining code, while also supporting personal growth by enhancing confidence and independence. However, developers reported significant challenges, including hallucinations and incomplete responses, limited knowledge for novel problems, struggles with logical reasoning, and concerns about impeding junior developer learning. These findings reveal a landscape where productivity gains must be balanced against quality assurance and educational concerns.

\item \textbf{Process -- RQ2: How have LLMs influenced software development processes?}

We mapped participants’ accounts onto the full software development lifecycle (SDLC), from requirements through maintenance. We found that LLMs have minimal impact on requirements gathering but are increasingly used for design ideation and problem decomposition. Implementation practices center on sophisticated prompt engineering strategies, where developers vary the specificity of prompts, use follow-up queries, and carefully manage security through generalized prompts. Code evaluation involves systematic verification through reading, output checking, and manual testing, with developers modifying code before integration and frequently discarding generated content. Developers turned to LLMs for generating unit tests and debugging, yet found them less reliable for complex tests, code review, and context window limitations. These findings demonstrate that LLM integration is highly phase-dependent, requiring developers to develop specialized strategies for different SDLC stages.


\item \textbf{Product -- RQ3: How has the use of LLMs influenced the software products created?}

We investigated how developers perceive the impact of LLMs on the code and software products they create, focusing on accuracy, readability, complexity, and security. Participants generally viewed LLM-generated code as clean and readable for common, small tasks but raised concerns about over-engineered solutions, outdated information, and potential vulnerabilities. Although the majority of participants considered LLM code sufficiently secure for their non-production use cases, many expressed concerns about sending confidential data to LLMs and the need for careful review before using generated code in production. These findings highlight a quality paradox: LLMs can produce acceptable code for specific contexts but require substantial developer oversight to ensure data security and production readiness. 
\item \textbf{Society -- RQ4: How may the software industry and education be affected by LLMs?}

We explored developers' perceptions of LLMs' impact on the software industry and educational training. We found that developers are actively redefining LLMs' organizational roles, comparing them to pair-programmers or junior developers, with some developers believing the overall job market will remain stable despite transformation. Widespread peer adoption coexists with concerns about entry-level positions and the need for organizational guidelines. In education, many participants advocated for integrating LLMs as learning tools, emphasizing the importance of teaching prompt engineering, while several stressed the continued importance of fundamental concepts and the need to adapt assignments and teaching methods. These views depict a profession in transition, grappling with how to use LLM capabilities while preserving core professional competencies and addressing equity concerns.


\end{itemize}

Our paper makes the following contributions:
\begin{itemize}
   
\item We provide a qualitative analysis of developers’ experiences with LLM-based tools across the software development lifecycle, mapping reported impacts from requirements through maintenance and identifying phase-specific patterns of adoption and challenge.

\item We characterize the strategies developers use to integrate these tools into their workflows, including prompt engineering techniques, code verification practices, and context management approaches that can inform tool design and organizational guidelines.

\item We treat our 2023 interviews as early empirical evidence from early adopters and, in the discussion, compare these accounts with more recent empirical and conceptual work on LLMs in software engineering, clarifying where early patterns have remained stable, or evolved in the current literature. 


\item We contribute empirical insights into the human factors and broader consequences of LLM adoption, including effects on developer confidence, learning, job roles, and educational practices, that complement technical evaluations and are critical for responsible use of LLMs in software engineering.

\end{itemize}
\section{Related Works}
The rapid integration of Large Language Models (LLMs) into software engineering practices necessitates a critical examination of their multifaceted impact. This paper provides such an examination across four key dimensions: the people (developers and teams), the processes (the development life cycle), the products (the software artifacts), and the broader societal implications. Due to thematic overlap, we have combined literature on the process and product subsections into a single section. The body of research in this domain is emerging at a significant pace. Although several systematic literature reviews have begun to map the field, such as recent work by Hou et al. on general LLM applications in SE \cite{hou_large_2024} and Ferino et al. on novice developer perspectives \cite{ferino_novice_2025}, the sheer number of publications (both peer-reviewed and preprints) makes a comprehensive survey challenging. This aligns with observations by Fan et al. \cite{fan2023large} that formal surveys struggle to capture all relevant, contemporary work. To reflect the current state of the field, we therefore include a small number of recent, formative non-peer-reviewed studies alongside peer-reviewed work.

\subsection{People}
\subsubsection{Developer Perceptions and Well-being}
Beyond broad systematic mappings, a growing body of qualitative work has begun to investigate developer-centric perspectives. Several interview studies of early adoption consistently report a tension between perceived productivity gains and concerns about tool reliability. For instance, Klemmer et al. found that while developers used AI assistants for security-critical tasks, they fundamentally lacked trust in the tool's output and felt compelled to double-check its work \cite{klemmer2024using}. This finding is mirrored in work by Rabani et al., who also noted developers' emphasis on LLM inaccuracy and the subsequent need for manual debugging \cite{rabani2023developers}. Similarly, Mendes et al. identified key challenges, including poor accuracy and the potential for distraction, which tempered the reported benefits of faster development \cite{mendes2024you}. While these studies capture developer perceptions, other work is beginning to quantify the impact of adoption on developer well-being. Feng et al., using a mixed-methods approach, modeled the relationship between Generative AI (GenAI) adoption and developer burnout, finding that GenAI adoption can heighten job demands and increase burnout, unless supported by positive attitudes and organizational resources \cite{feng_burnout_2025}.

\subsubsection{Novice vs. Senior Developers}
The ``People'' dimension of LLM adoption is further nuanced by developer experience, which strongly shapes the utility and risk of these tools. A body of work has concentrated on novice developers, for whom LLMs present a distinct dual-use scenario. A study by Kazemitabaar et al. indicates that while novices value LLMs for learning support, they concurrently struggle with over-reliance and face challenges in accurately evaluating AI-generated code \cite{kazetimbar_novices_2024}. This is reinforced by Siddiq et al., who observed that novices may treat LLMs as a ``black box," potentially impeding the development of fundamental skills \cite{siddiq2024quality}. In contrast, senior developers tend to employ LLMs as a ``junior partner'' for task augmentation, using them to generate boilerplate code or alternative implementations while relying on their own expertise to vet and refine the output \cite{FRANCE2024649}.

\subsubsection{Team Dynamics and Roles}
Beyond the impact on individual developers, emerging research suggests that LLM integration is beginning to reconfigure team dynamics and collaborative processes. The acceleration of code generation, for instance, can introduce new pressures and bottlenecks in subsequent stages of the ``Process,'' particularly code review and quality assurance \cite{banh_genaiswe_2025}. The ``People'' dimension is also highly role-dependent. Thapa et al., investigating the specific impact on software testers, found that LLMs are viewed with ambiguity compared to engineers: they are seen as valuable tools for augmenting test case generation but also as a potential threat that could devalue the human-centric and exploratory aspects of quality assurance \cite{thapa_css_2025}.

\subsection{Process and Product}
\subsubsection{Empirical Evaluation of LLMs in the SDLC}
Moving from developer perceptions to empirical performance, another research vector has focused on evaluating LLMs against specific software engineering tasks. A primary focus has been on productivity and code generation. Peng et al., for example, conducted a controlled experiment observing a significant positive impact on developer productivity, with the AI-assisted group completing tasks more quickly \cite{peng2023impact}. Beyond general productivity, a growing number of studies are empirically assessing LLM capabilities in core ``Process'' and ``Product'' activities, particularly code refinement and quality assurance. For example, Guo et al. conducted an empirical study on the potential of ChatGPT for automated code refinement based on code reviews, finding that it outperformed a state-of-the-art baseline \cite{guo2024exploring}. Similarly, Wadhwa et al. presented CORE, a tool utilizing an LLM-based proposer-ranker architecture to successfully resolve static code analysis warnings, demonstrating a practical application for improving code quality \cite{wadhwa2024core}. However, human oversight in these processes remains critical; both Alomar et al. and Siddiq et al. observed that although developers using AI-based code review tools were more efficient, they were also more likely to accept flawed code, highlighting a new set of risks for the ``Process'' dimension \cite{alomar2024refactor, siddiq2024quality}.

\subsubsection{Testing and Debugging}
A significant sub-field of this empirical investigation concerns the application of LLMs to software testing and debugging—a core ``Process'' that directly governs ``Product'' quality. Zhang et al. present a comprehensive systematic literature review that concludes LLMs could automate the generation of a majority of unit tests and even identify potential bugs, albeit with varied performance \cite{zhang_unitests_2025}. Tanaka et al. further explored this by using LLMs to find and repair broken test cases, achieving a notable success rate \cite{tanaka_gentests_2025}. More recently, research has focused on highly specific and challenging testing issues. Fatima et al., in their work on FlakyFix, successfully used LLMs not only to predict the category of fix required for a flaky test but also to perform the test code repair itself, demonstrating a guided repair process that leverages LLM capabilities \cite{fatima2024flakyfix}. These studies suggest a strong potential for LLMs to augment quality assurance workflows, although the reliability and scalability of these techniques are active areas of investigation.

\subsubsection{Security Risks and Mitigation}
The implications of LLMs for the software ``Product" extend critically into the domain of security, presenting a significant dual-use challenge. On one hand, LLMs are being leveraged as a component of ``Process" to improve security; for example, by powering advanced fuzzing frameworks \cite{sun_fuzzing_2024} designed to detect vulnerabilities in complex contexts. On the other hand, a substantial body of work demonstrates that LLMs can inadvertently introduce critical vulnerabilities \cite{islam_vuln_2024, gabison_liability_2025}. This is compounded by the ``self-deceive" problem, where an LLM's programmatic goal to be helpful results in the generation of insecure code that developers, who may not be security experts, might overlook \cite{ barkur_deception_2025}. This risk has prompted a reactive line of research focused on mitigation, resulting in frameworks such as SafeGenBench, which aim to detect and mitigate security-related risks inherent in LLM-generated code \cite{li_safegenbench_2025}.

\subsubsection{Code Quality and Innovation}
The integration of LLMs into the development ``Process" has a direct, albeit complex, effect on the software ``Product." A primary area of concern is code quality. While LLMs can accelerate the quantity of code produced, a significant body of research, as previously discussed, has raised concerns about the quality and security of that output \cite{li_safegenbench_2025, gabison_liability_2025, islam_vuln_2024}. Beyond mere code generation, however, LLMs are also being positioned as tools for innovation and ``Product" enhancement. For example, researchers are exploring their use in generating novel UI/UX designs and even in drafting initial software documentation, thereby expanding the scope of what constitutes the ``Product" \cite{chen_genui_2025}. This suggests the impact on the ``Product" is twofold: a high-volume, high-risk impact on code artifacts and a high-potential, nascent impact on innovation and non-code artifacts.

\subsection{Society}
\subsubsection{Legal and Ethical Implications}
Beyond the technical dimensions of 'People,' 'Process,' and 'Product,' the adoption of LLMs introduces profound 'Societal' implications, particularly regarding legal and ethical frameworks. The ambiguity surrounding the provenance of LLM training data has initiated a critical debate among legal scholars about copyright, data privacy, and intellectual property attribution \cite{francesconi_linkedopendata_2023}. These concerns are not merely theoretical; they have direct consequences for software development. Gabison et al. examined the practical legal challenges of liability for agentic agents and their outputs \cite{gabison_liability_2025}. Furthermore, empirical studies such as LiCoEval have shown that LLM outputs may be subject to existing open-source licenses, creating complex and unresolved compliance issues for organizations that adopt these tools \cite{xu_licoeval_2025}.

\subsubsection{Impact on Computing Education}
The integration of LLMs into pedagogical settings is being actively investigated to understand their effect on student learning and skill development. For example, Tanay et al. observed that upper-level computing students using LLMs for software engineering projects demonstrated improved efficiency in task completion and information retrieval \cite{tanay2024exploratory}. Other work by Essel et al. similarly found that tools like ChatGPT could enhance undergraduate students' critical, reflective, and creative thinking skills \cite{essel2024chatgpt}. This emergent body of evidence has catalyzed discussions on curriculum adaptation. Ozkaya, for instance, has proposed that curricula must evolve to teach students how to collaborate effectively with AI, particularly for maintaining legacy systems \cite{ozkaya2023application}. Further supporting this, Jeuring et al. identified a strong correlation between computational thinking (CT) skills and a student's ability to co-develop effectively with ChatGPT, reinforcing the continued necessity of teaching fundamental CT skills \cite{jeuring2023skills}.

\subsubsection{Literature Gap in Practitioner Perceived Effects on Education}
While these studies provide valuable initial insights into how LLMs may augment student learning from an academic perspective, they illuminate a clear gap in the literature. There is, to date, limited research focused on the perspectives of professional developers regarding the role of LLMs in reshaping computing education. The views of active industry practitioners are a critical, missing component in the discussion of how to align educational practices with new, AI-driven professional settings. Our study helps to address this gap by engaging developers directly, capturing their experiences and views on how LLMs will influence the future teaching and learning of computing skills.

\subsubsection{Perceptions and Risks}
The ``Societal" dimension of LLM adoption is characterized by a significant ``double-edged sword" narrative, where perceived benefits are weighed against substantial risks. This duality is evident in public and developer discourse. On one hand, LLMs are seen as tools that can lower the barrier to entry for programming \cite{guzdial_scaffolding_2023}. On the other hand, this same accessibility is a source of societal risk; several corporate threat assessment groups such as those at Google and Microsoft have begun reporting the use of LLMs for malicious purposes, such as generating malware or exploiting vulnerabilities at scale \cite{google_threat_2025, microsoft_threat_2025}. This tension creates a complex ``Societal" landscape where the technology is simultaneously viewed as a catalyst for innovation and a potential vector for significant harm.

\subsubsection{Job Market and Displacement Anxiety}
A dominant societal theme arising from widespread LLM adoption is the profound anxiety surrounding the future of the software engineering job market. The popular press and, to some extent, the academic literature have amplified concerns that LLMs could lead to widespread job displacement, de-skilling, or the devaluation of developer expertise \cite{ferino_novice_2025, hou_large_2024}. This narrative of automation-induced obsolescence is a frequent topic of investigation. However, research focusing on developer perceptions, including foundational work by Klemmer et al. and Mendes et al., suggests a more nuanced reality: developers often view these tools not as replacements, but as augmentations that shift, rather than eliminate, the need for human expertise \cite{klemmer2024using, mendes2024you}. This perspective reframes the "Societal" impact as an evolution of the developer role rather than its extinction.

Taken together, existing work offers rich but fragmented insights into how LLMs affect software engineering. Prior studies typically focus on specific roles (such as students, novices, testers, or individual developers), particular activities (such as code generation, testing, or security), or single dimensions of impact. There is still limited qualitative work that follows professional early adopters across the full software development lifecycle and situates their experiences within a broader sociotechnical framework spanning people, process, product, and society. Our study addresses this gap by providing an in-depth view of early LLM adoption in professional software development and by comparing these accounts with subsequent empirical and conceptual work to show how early patterns have persisted, or shifted.

\section{Methods}
\subsection{Participants and Demographics}
We recruited 16 participants from technology meetup groups in the Southeastern United States, as well as through the researchers' personal contacts and LinkedIn. Snowball sampling was also employed, where participants invited individuals from their networks to join the study \cite{sadler_research_2010}.
The target population for this study was professional software engineers and developers. Inclusion criteria required participants to: (1) be employed as full-time professional software engineers or developers; (2) have begun using ChatGPT or similar LLM-based tools between November 2022 and April 2023 (qualifying them as early adopters); and (3) have at least two months of active experience using these tools for programming tasks in their professional work by the time of the interview. Initially, our call for participants attracted a number of graduate students. Although many had prior experience as developers or had worked in development roles at their universities, we determined that this group would not fully address our research questions. As a result, we revised our recruitment materials to focus specifically on full-time developers. The demographic details of all 16 participants are provided in Table \ref{table:demographics}. Our sample size of 16 participants is consistent with recommendations for 
qualitative interview studies seeking thematic saturation, where Guest et al. found that 12-15 interviews typically suffice to identify the majority of themes in relatively homogeneous populations \cite{guest2006many}. Our sample size enabled us to reach thematic saturation across our four research dimensions while maintaining the depth of analysis expected in qualitative inquiry.

\subsection{Interview Questions Formulation}
We structured the interview protocol around four key themes in software engineering: People, Processes, Products, and Society. These themes were derived from prior work on LLMs in software engineering and sociotechnical views of software development, which emphasize people, processes, products, and societal context as critical dimensions. Using this foundational framework, we formulated interview questions designed to capture relevant insights within each theme. We developed the questions iteratively to ensure their clarity, relevance, and alignment with our research questions.

To refine and ground our interview approach, we conducted several pilot interviews. One pilot interview was with a professional developer who regularly uses ChatGPT for programming, and the other three were with students who self-identified as developers and used ChatGPT in their work, though they were not employed as full-time professionals.  These pilot interviews served as a testing phase to assess the appropriateness of the interview questions, determine the time required for each interview, and evaluate the overall structure. While the student interviews were excluded from our analysis, the professional developer’s interview was included due to the rich data it provided.

Following the pilot interviews, we reviewed the data and feedback, ultimately excluding the student interviews from the preliminary analysis due to concerns about their alignment with the study's objectives. Based on the insights gained, we refined the questions, reworded them for clarity, checked for grammatical accuracy, and ensured the wording was broadly accessible to participants with different professional backgrounds. The final version of the questions used for the interviews can be found in Table \ref{Interviews}. Additionally, we reduced the interview duration to ensure the questions could be answered efficiently without sacrificing the depth of the responses.

\subsection{Procedure}

\subsubsection{Interview}
\begin{table}[]
\caption{A list of the interview questions we asked participants.}
\begin{tabular}{|l|}
\hline
How do you use ChatGPT in your everyday work?                                                                                                                                        \\ \hline
How often (on average) – multiple times a day? Daily? Weekly? Monthly?                                                                                                               \\ \hline
Can you tell me about a time when you used ChatGPT to help you write a program?                                                                                                      \\ \hline
How about a time when it did not work out well?                                                                                                                                      \\ \hline
Were there any times when you were surprised by what you could do with ChatGPT and code?                                                                                             \\ \hline
How has ChatGPT changed your software engineering process?                                                                                                                           \\ \hline
Has it changed how you gather requirements? How?                                                                                                                                     \\ \hline
Has it changed how you break tasks into parts that can be solved by ChatGPT? How?                                                                                                    \\ \hline
Has it changed how you write code? How?                                                                                                                                              \\ \hline
Has it changed your testing process? How?                                                                                                                                            \\ \hline
Has it changed your code review process? How?                                                                                                                                        \\ \hline
How do you normally evaluate the code generated by ChatGPT?                                                                                                                          \\ \hline
How do you do testing?                                                                                                                                                               \\ \hline
\begin{tabular}[c]{@{}l@{}}How do you determine if the code does what you asked for? \\ Do you read the code?\end{tabular}                                                           \\ \hline
\begin{tabular}[c]{@{}l@{}}How do you check the code quality, efficiency, complexity? \\ What about security aspects?\end{tabular}                                                   \\ \hline
How much do you trust the code provided by ChatGPT?                                                                                                                                  \\ \hline
How is it different from evaluating human-written code?                                                                                                                              \\ \hline
How do you integrate the ChatGPT code results into your codebase, if at all?                                                                                                         \\ \hline
\begin{tabular}[c]{@{}l@{}}What steps do you take before integrating the output of ChatGPT into your code? \\ Or adapting/modifying the ChatGPT code to make it useful?\end{tabular} \\ \hline
How often do you throw away the code, use, or reuse the code given by ChatGPT?                                                                                                       \\ \hline
How secure do you believe the code given by ChatGPT generally is?                                                                                                                    \\ \hline
How will ChatGPT impact the skills and jobs in the software industry?                                                                                                                \\ \hline
How do you think CS degree programs should adjust to prepare for this shift?                                                                                                         \\ \hline
What skills must a person have to use ChatGPT like you do?                                                                                                                           \\ \hline
For example, how do you formulate a question to ChatGPT?                                                                                                                             \\ \hline
How do you structure your queries to get the desired answer?                                                                                                                         \\ \hline
How broad or specific should your questions be?                                                                                                                                      \\ \hline
How much context will you provide to ChatGPT?                                                                                                                                        \\ \hline
How do follow-up questions impact the accuracy of your answer?                                                                                                            \\ \hline
How many queries/reformulations                                                                                                                                                      \\ \hline
\begin{tabular}[c]{@{}l@{}}Do you or your company have any guidelines, formal or informal, \\ about how developers should use ChatGPT?\end{tabular}                                  \\ \hline
\end{tabular}
\label{Interviews}
\end{table}

We invited interested participants via email and asked them to complete consent forms. Participants used a Google Calendar link to schedule their own research sessions, which lasted approximately 70 minutes. All interviews took place between March 1, 2023, and July 7, 2023. While all participants consented to audio recording, not all agreed to the screen or video capture. Audio recordings provided sufficient data for our analysis, as our focus was on participants' experiences and perspectives rather than detailed observation of their interactions with tools. 

The study consisted of semi-structured interviews, featuring a core set of predetermined questions, while allowing flexibility for follow-up questions based on the conversation \cite{miles_handbook_2005}. At the start of each session, participants were briefed on the research purpose (approximately 10 minutes), followed by the semi-structured interview (approximately 60 minutes). Two researchers were present: one conducted the interview, while the other took notes. Both researchers asked clarifying or follow-up questions based on participant responses.

During the interviews, participants were asked to discuss their use of LLMs in their daily work, how they evaluated LLM-generated code, and how they integrated it into existing codebases. We also explored their opinions on how LLMs might impact the skills and job landscape in the software industry, with specific follow-up questions regarding the skills needed to effectively use LLMs. Additionally, participants were asked to describe real projects they had worked on. While the questions we asked explicitly named ChatGPT, we stated during the interviews that interviewees should also answer the questions based on their usage of similar LLM-based chatbots and other LLM-based tools.

\subsubsection{Demographic Survey}
\begin{table}[]
\caption{Demographics of participants P1-P16 (P\#) showing the company size (number of employees), gender, race/ethnicity, years of development experience, and LLM tool(s) used.}
\begin{tabular}{|l|l|l|l|l|l|l|l|}
\hline
\textbf{P\#} & \textbf{Company size} & \textbf{Gender}     & \textbf{Trans} & 
\textbf{Race / ethnicity}            &  \textbf{Experience} & \textbf{LLM used}         \\ \hline
P1           & Small (10-19)        & Non-binary & No    
& Hispanic, White                     & 3+ years         & GPT               \\ \hline
P2           & Medium (50-249)       & Man        & No    
& Asian                     & 1-2 years        & GPT               \\ \hline
P3           & Large enterprise (250+)         & Man        & No    
& Hispanic, Asian                     & 3+ years         & GPT               \\ \hline
P4           & Micro (5-9)          & Man        & No    
& Hispanic, White                     & 3+ years         & GPT               \\ \hline
P5           & Large enterprise (250+)         & Man        & No    
& White                     & 3+ years         & GPT               \\ \hline
P6           & Micro (1-4) & Man        & No    
& Black or African American & 3+ years         & Copilot          \\ \hline
P7           & Large enterprise (250+)         & Man        & No    
& Asian                     & 3+ years         & GPT               \\ \hline
P8           & Medium (50-249)       & Man        & No    
& Asian                     & 3+ years         & GPT               \\ \hline
P9           & Large enterprise (250+         & Man        & Yes   
& White                     & 3+ years         & GPT               \\ \hline
P10          & Small (20-49)        & Man        & No    
& Asian                     & 3+ years         & GPT               \\ \hline
P11          & Micro (1-4) & Man        & No    
& Asian                     & 3+ years         & GPT               \\ \hline
P12          & Large enterprise (250+)         & Non-binary & Yes   
& White                     & 3+ years         & Bard                  \\ \hline
P13          & Large enterprise (250+)         & Woman      & No    
& Asian                     & 1-2 years        & GPT               \\ \hline
P14          & Micro (5-9)          & Man        & No    
& White                     & 1-2 years        & GPT               \\ \hline
P15          & Medium (50-249)       & Man        & No    
& White                     & 3+ years         & GPT, Copilot 
\\ \hline
P16          & Large enterprise (250+)         & Woman      & No    
& Black or African American & 1-2 years        & GPT               \\ \hline
\end{tabular}

\label{table:demographics}
\end{table}

After completing the interview, participants were asked to complete a brief survey on Qualtrics, which collected demographic information. Specifically, participants were asked to provide details about the size of their company, as well as their gender, ethnicity, and race. This demographic data was gathered to explore potential differences in LLM usage across various groups. Based on the information provided during the interviews, we also estimated each participant’s years of experience in software development. These results are presented in Table \ref{table:demographics}. Our participants were 43.75\% White, 43.75\% Asian, and 12.5\% Black. 75\% were men, 12.5\% were women, and 12.5\% were non-binary. 75\% of our participants had at least 3 years of development experience, with only four participants having more limited experience and two participants expressing they had begun their careers within the past year.

\subsection{Analysis}

All interviews were recorded, automatically transcribed, and then reviewed and corrected by the researchers, who re-watched the video recordings to ensure accuracy.

Our analysis followed an inductive approach, similar to that of Silva et al. in their study of AR activists \cite{silva_understanding_2022}. Initially, we created open codes based on three transcripts that we identified as thematically rich and highly relevant to our research questions. After discussing and refining these codes, we compiled a preliminary codebook with descriptions for each tag. The two researchers who conducted the interviews, along with one researcher who did not participate in the interviews (providing an independent perspective without potential interview bias), independently coded the remaining 
13 transcripts. Following independent coding, the researchers held discussions to reconcile their codes. All tags included in the final codebook required agreement from at least two of the three coders, with disagreements resolved through discussion until consensus was reached. This collaborative approach ensured coding consistency while preserving the emergent, inductive nature of our analysis and allowed us to leverage multiple perspectives in interpreting participant experiences.

Subsequently, three researchers (two involved in coding and one external) grouped the tags according to each research question and then clustered them into mid-level themes. These were further divided into lower-level themes, which served as the foundation for our written analysis. This process took approximately two weeks, consisting of multiple intensive coding sessions. Throughout the analysis, the researchers continued to refine the codes and themes through ongoing discussions and by sharing draft results. The final coding resulted in 361 total codes, categorized into 138 low-level themes. For each theme, we also quantified the number of quotes tagged, which we present in our results. 
Participant quotes are referenced as PX, where X denotes the participant's interview order. 

To ensure analytic rigor, we employed several strategies to mitigate researcher bias. First, we included one researcher in the coding process who had not conducted interviews, providing an external perspective. Second, we maintained detailed audit trails of our coding decisions and theme development. Third, we grounded our interpretations in direct participant quotes and quantified the prevalence of themes. Finally, we remained reflexive about our own positions as researchers in academia, studying professional practice, and carefully distinguished between participants' perspectives and our interpretations.

\section{Results}
\label{results}
In this section, we present findings from our thematic analysis of 16 interviews with early-adopter software professionals. We organize our results around four research questions that comprehensively address the sociotechnical dimensions of LLM integration: how LLMs affect developers (RQ1), influence development processes (RQ2), impact software products (RQ3), and shape broader industry and educational contexts (RQ4). For each research question, we present both benefits and challenges identified by participants, with theme prevalence indicated by the number of interviews in which each theme emerged. All findings are grounded in participant experiences and illustrated with representative quotes.

\subsection{RQ1: How do LLMs affect developers?}
The key themes that emerged in understanding how LLMs affect developers' daily tasks are as follows:
\subsubsection{\textbf{Boosting Developers' Productivity}}
\begin{itemize}
\item \textbf{Reducing Mundane Tasks:} Codes related to LLMs' effectiveness in simplifying mundane tasks appeared in fourteen interviews, aligned with previous research describing how developers used ChatGPT to automate tedious tasks \cite{hornemalm_chatgpt_2023}. A notable example was provided by P14, who shared, ``\textit{I really like that the repetitive, boring tasks, like looking for a comma; that I don't need to do those, and I can focus more on building things.}''
Thirteen participants highlighted LLMs' role in expediting the software engineering process and saving time, while nine participants emphasized how LLMs enhance efficiency and make developers more productive. For example, P15 explained how ChatGPT writes functions, supporting focus on higher-level tasks: ``\textit{[Suppose that] I want to do something simple [...] I don't write those functions anymore. I always have ChatGPT do it because I know that it's going to come up with something close to what I was going to do, but I didn't actually have to do it. So it's kind of like a pair-programmer for me, so I can stick to the higher-level stuff that would take more understanding of the infrastructure.}''
\item \textbf{Streamlining Search Experiences:} Ten participants highlighted the productivity advantage of LLMs in reducing search time for solutions. For example, P14 mentioned how LLMs transformed his search style: ``\textit{Now, I haven't been on Google probably for 3 weeks [...] I don't need to click on each of the links from Google to find the solution[...] [ChatGPT] gives me a summarized version from possibly three, four different sources.}''

\item \textbf{Providing Boilerplate Code and Templates:} Ten participants shared that they had used LLMs to create generic or standardized code and templates. P4 described this capability as ``\textit{one of the best things I get from [ChatGPT].}'' P12 expressed, ``\textit{You can take [Bard's response] as an idea, like a first draft, and then iterate on it; that's the experience I have working with it from the coding side.}''

\item \textbf{Additional Productivity Benefits:} Code translation was referenced in three interviews. Two participants noted that ChatGPT reduces the learning curve and speeds up learning. Two participants mentioned using LLMs for set-ups and installations. One participant highlighted how ChatGPT could supplement incomplete or outdated tutorials and documentation, while another noted its usefulness in improving recall of syntax and implementation details.
\end{itemize}
\paragraph{\textbf{Challenges:}}
\begin{itemize}
\item \textbf{Programming Language Mix-Ups:} Three participants noted instances where LLMs' responses were not in the expected language, requiring developers to spend time converting the code to the target language.
\item \textbf{Other Challenges:} One participant observed that ChatGPT's answers sometimes contradicted its earlier responses. One participant noted slow response generation with GPT-4 impacted user experience. One participant discussed challenges with unstructured data analysis.
\end{itemize}
\subsubsection{\textbf{Facilitating Developers' Learning}}
\label{sec:devlearning}
\begin{itemize}
\item \textbf{Personalizing Education and Skill Enhancement:} Learning new technologies, software, and information from LLMs emerged as a prominent theme, appearing in fourteen interviews. A key benefit of LLMs is their ability to offer personalized and interactive learning. For example, P7 highlighted the constant need to learn new things in the industry: ``\textit{We still need to read a lot of things like documentations or ...some new technologies. So ChatGPT is a good resource...to learn something I never heard before. So, for example, if there is a question [and] I'm not sure which tool I should use, I could probably just ask the open question from ChatGPT.}''
\item \textbf{Explaining Code:} LLMs' abilities to provide explanations and examples were noted in 11 interviews. P13 shared how ChatGPT was more useful than other tools due to its ability to explain code more efficiently: ``\textit{[My] first step tends to be to go to ChatGPT, give it the snippet of my code, [...] and then ask it those specific questions because I think it puts you quite ahead in understanding and making progress with things.}'' 

\item \textbf{Providing a Broad Scope of Knowledge and Diverse Datasets:} The extensive knowledge and diverse datasets of LLMs appeared in four interviews as an element that advanced developers' knowledge and understanding. 

\item \textbf{Demonstrating Best Coding Practices:} Two participants highlighted how LLMs helped them learn how to write code more efficiently and elegantly by observing the generated code and incorporating LLMs' suggestions.
\end{itemize}
\paragraph{\textbf{Challenges}}
\begin{itemize}
\item \textbf{Hallucinations and Incomplete Responses: } This challenge surfaced in six interviews. For instance, P1 recounted instances where ChatGPT fabricated responses: \textit{There were some things that were surprising in a bad way, like it would make up papers. It would hallucinate paper names and authors.}'' Another participant, P11, highlighted ChatGPT's failure to provide sources for its responses, noting the importance of human expertise: \textit{[People's] knowledge is probably more up to date [...] ChatGPT really does not cite its information. I think maybe Bard started citing stuff. Citations are very helpful.}''
\item \textbf{Limited Knowledge and Datasets:} Five participants shared that LLMs have limited knowledge and datasets. For instance, P9 shared that despite extensive efforts and careful prompt engineering, ChatGPT failed to provide an answer, requiring him to hire a JavaScript subcontractor.

\item \textbf{Struggles with Novel Ideas and Logical Prompts:} In four interviews, participants highlighted LLMs' limitations in generating code for novel ideas and handling prompts requiring complex logical reasoning.

\item \textbf{Impediment of Developers' Learning:} Four interviews raised concerns about LLMs' probable adverse effect on learning, particularly for junior developers, primarily due to users' potential inability to parse the correctness of ChatGPT's answers.

\item \textbf{Safeguard Failures and Harmful Effects: } In two interviews, participants reported incidents where LLMs' safeguards against disseminating harmful information could be bypassed.
\end{itemize}
\subsubsection{\textbf{Supporting Developers' Personal Growth}}
\begin{itemize}
\item \textbf{Enhancing Reassurance, Confidence, and Independence:} This theme emerged in six interviews. P2, a newer professional, shared his experience of comparing his answers to ChatGPT's for corroboration: ``\textit{It has made the [software engineering] process easier definitely. And the best part is I kind of get the reassurance of a lot of things --- that if I'm doing it this way, I can ask it, and I can just check if ChatGPT is giving me a similar answer [...], then I'm not doing anything which is majorly wrong.}''
\item \textbf{Improving Access to Information:} A positive view of LLMs' ability to ease access to information emerged in three interviews, with participants believing ChatGPT would democratize information access worldwide.

\item \textbf{Enhancing Job Satisfaction:} One participant highlighted ChatGPT's role in his increased job satisfaction by helping him with tasks he struggled with, allowing him to focus on what he cared about.
\end{itemize}
\paragraph{\textbf{Challenges}}
\begin{itemize}
\item \textbf{Inability to Replace Human Decisions}: Six participants noted LLMs' inability to replace human interactions and decisions. For instance, P12 emphasized that software engineering involves more than just coding: ``\textit{Despite what [LLMs] can do, at the end of the day, it can't replace --- like, what's difficult about being a software engineer isn't coding particularly. [...] I think what's difficult about being a software engineer, at least with what I do, is the communication that happens between teams, between coworkers, the email threads, and the chat threads that exist.}''
\item \textbf{Concerns about Dependency:} Two participants mentioned their preference for solving problems on their own before asking LLMs to avoid over-reliance.

\item \textbf{Slower Implementation than Humans:} One interview noted differences between the time it took humans and LLMs to create code in cases where ChatGPT fails to understand or provide accurate solutions.
\end{itemize}
\subsubsection{\textbf{Assisting Developers' Non-Technical Tasks}}
\begin{itemize}
\item \textbf{Consulting and Decision-Making:} Eleven participants discussed the use of LLMs for consultation or direction. For example, P8 mentioned ChatGPT's impact on resolving team debates: ``\textit{[A]lways in technical teams, there exist debates on choosing options, options A and B -- both are correct, but which one is better? In this way, we have a judge; we have someone that tells us which approach is better.}''
\item \textbf{Summarizing Text and Documentation:} Nine participants mentioned the use of LLMs for summarizing different textual data, such as articles, papers, and documentation. 

\item \textbf{Supporting Internal Communication:} The use of LLMs to facilitate internal communication by composing documents, aiding in presentations, and assisting in explaining complex tasks to team members emerged in five interviews.
\end{itemize}
\paragraph{\textbf{Challenges}}
\begin{itemize}
\item \textbf{Limited Summarizing and Explanation Capabilities:} Two participants expressed dissatisfaction with LLMs' ability to summarize and explain, with one noting ChatGPT's tendency to make generalizations without actually reading provided materials.
\end{itemize}


\subsection{RQ2: How have LLMs influenced software development processes?}
To answer RQ2, we loosely organize our findings based on the software development life cycle \cite{AWSndsdlc} and the agile development methodology \cite{fowler2001agile}. We identified impacts across five key phases: (1) Requirements and Planning, where LLM adoption was minimal; (2) Design and Ideation, where LLMs encouraged problem decomposition; (3) Implementation, where developers employed sophisticated prompt engineering strategies; (4) Testing and Code Review, where LLMs showed promise for unit tests but faced limitations; and (5) Debugging, Refactoring, and Documentation, where LLMs reduced debugging time. We organize our findings to follow this progression through the development lifecycle.

\subsubsection{\textbf{Requirements and Planning}} \label{sec:requirements} We use a common definition of requirements, defining it as a software capability that must be met by a system or system component in order to satisfy a specification \cite{leffingwell2000managing}. Similarly, we define planning as the process of collecting requirements from stakeholders, scheduling, and resource estimation/allocation \cite{AWSndsdlc}.
\begin{itemize}
\item \textbf{Discovering Missing Components:} Two participants utilized LLMs for uncovering missing components, as P15 shared: ``\textit{I have an idea of what I think it should be, [...] and then I put it in ChatGPT and say, `What are some gotchas, or what are some things that I'm missing? Or what kind of questions should I be asking in order to fill in any blanks that I might have?' So [ChatGPT is] kind of my consultant, as if it was a more senior developer than me, or more a manager than me, or something like that, to where I'm going to get some feedback.}''
\item \textbf{Prototyping and Refining Requirements:} Two participants found LLMs proficient in prototyping, and two found them helpful in refining requirements, especially for independent contractors. P9 shared how ChatGPT helped him research and cut down discovery time when interacting with stakeholders: ``\textit{Say, if I got a new contract job that I'm scoping to try to help my clients to try to refine the technical requirements, and especially so on the domains I'm not familiar with [...] while I was talking to them getting requirements at the same time, I would have the ChatGPT window on the side.}''
\end{itemize}
\paragraph{\textbf{Challenges}}
\begin{itemize}
\item \textbf{Inability to Replace Human Involvement:} Twelve participants stated that LLMs did not have an impact on their requirements gathering. Participants mentioned that they received requirements and plans from their supervisors or superiors, who work directly with client needs, and these can't be generated by LLMs. As P2 shared, ``\textit{[Requirements gathering] hasn't changed because most of the work that I'm doing is with client developers and product managers from the client side. So we need to talk to them for the requirements we need to specify [...] And because our projects are very client-specific, I cannot just go on the Internet to see what they would expect.}''
\item \textbf{Limited Requirements Detailing:} One participant shared how ChatGPT cannot help to generate detailed requirements, noting that responses were ``\textit{very common sense}'' but didn't address usability aspects or fine-tuned details. Another participant noted struggles with multi-criteria decision-making.
\end{itemize}
\subsubsection{\textbf{Design and Ideation}}
We consider design to be the process of finding solutions and creating a more detailed technical plan as a result of the requirements-finding process \cite{AWSndsdlc}, where ideation is a critical component in designing. Decomposition is also part of the design process, and Chattopadhyay et al. have noted how developers employ particular strategies in order to decompose their tasks into smaller units \cite{chattopadhyay2019latent}.
\begin{itemize}
\item \textbf{Increasing Problem Decomposition:} Eight participants acknowledged that LLMs encouraged or necessitated problem decomposition into smaller, more manageable components. This is because LLMs can only take in a limited amount of information at a given time. P4 shared that for him, ChatGPT ``\textit{might force me to break [problems] down [...] I guess it might be a good forcing function to make me break them down in the subtasks and subrequirements [...] If I can't explain it to ChatGPT, then it might be an indicator that I don't yet understand it.}''
\item \textbf{Utilizing for Ideation:} Six participants acknowledged using LLMs for ideation, leveraging their extensive knowledge bases and generative capabilities. P12 shared that for them, ``\textit{Probably the ideation stage is where [Bard] shines the most, because you can't trust it to be accurate or exactly what you want, but you can trust it to generate some cool content for you, some cool code ideas, some cool writing ideas.}''
\end{itemize}
\paragraph{\textbf{Challenges}}
\begin{itemize}
\item \textbf{Inapplicable for Decomposition:} Seven participants shared that LLMs have not affected their problem decomposition, viewing that as inherently reliant on individual reasoning. P13, for instance, stated, ``I don't think [it has impacted decomposition]. I like planning them myself and then going to ChatGPT for more fine-grained planning about how I should actually implement the code.''
\item \textbf{Stubborn Responses:} Six participants noted that LLMs were stubborn, which could limit their usefulness while designing. P11 remarked on ChatGPT's struggles to adapt to new input: ``\textit{[ChatGPT] doesn't do a good job of changing itself.}''
\end{itemize}
\subsubsection{\textbf{Implementation}} \label{sec:implementation}
The implementation process with LLMs parallels code reuse. Therefore, we adopted Rosson and Carroll's classification \cite{Rosson1996}, dividing the process into three distinct phases: (1) \emph{Finding Context,} which entails prompt engineering and the process of locating pertinent responses; (2) \emph{Evaluating Context,} which centers on evaluating the accuracy and usability of the generated responses; and (3) \emph{Integration,} which pertains to incorporating the generated text into an individual's own code or within an existing codebase. We note that the following bullet points' titles are framed from the perspectives of the developer/prompter, rather than the perspectives/capabilities of the LLMs.
\paragraph{(1) Finding Context}
\begin{itemize}
\item \textbf{Varying Prompt Specificity to Impact Answers.} Developers' diverse preferences for broad prompts, specific prompting, or adjusting query specificity to influence LLMs' output emerged in fifteen interviews.
Broad questions are favored when seeking varied answers or avoiding assumptions. P15 shared that while he originally had used specific prompts, he found that using broader prompts resulted in improved outcomes: ``\textit{[I]t started to give me actually better results, because it knew things that I didn't know [...] When I have a more broad [prompt], then it's able to kind of formulate its own deduction and get to the problem that it thinks that it's solving.}''
Specific prompts are preferred when seeking particular answers, especially in software engineering. As P2 stated, ``\textit{I think it should be very specific [...] If you are going into other [non-software] domains, where there is no one correct answer, you can go for broad questions. But for software development, I think you need to be very specific.}''
Four participants acknowledged that there is not a single correct strategy for choosing the level of specificity, emphasizing the importance of adaptability and experimentation.
\item \textbf{Enhancing Accuracy and Clarity with Follow-Up Queries:} Fifteen participants shared their strategies of prompting iteratively until getting the desired response. Seven participants shared that follow-up queries improve accuracy, and two shared that they improve clarity. As P16 stated, ``\textit{The follow-up questions actually might be more important than [the initial question], more probably equally important. But I get more of my answers from the follow-up question.}''
\item \textbf{Unique Prompting Strategies:} Fourteen participants acknowledged the crucial role of prompt engineering in eliciting accurate and relevant responses from LLMs. P9 shared that he could find a good answer ``\textit{[a]s long as the prompt is crafted carefully, and the problem is common enough to find a solution.}'' Participants demonstrated a willingness to try different prompting strategies, with inspiration from online resources and previous interactions with LLMs.
\item \textbf{Generalizing Prompts and Code for Security:} Thirteen participants employed generalized prompts or used case-specific information rather than project-specific details when interacting with LLMs. P2 shared that, ``\textit{Because I cannot put all the information into ChatGPT, because it's very client-specific information, it's confidential information [...] I just give it use cases that are similar.}''
\item \textbf{Providing Examples and Context:} Nine participants practiced few-shot prompting with ChatGPT, and nine participants highlighted the significance of providing context when interacting with LLMs. P1 observed that ``\textit{The more context, the better.}'' Two participants discussed how ChatGPT's ability to retain context distinguishes it from traditional search engines.
\item \textbf{Starting a New Thread for Fresh Answers:} Nine participants stated their preference for beginning a new thread when they were dissatisfied with the answers received in a previous interaction. P15 detailed his experience: ``\textit{[I] have it continue refining what I want to do. Sometimes, it'll get to the point where it kind of [stops] working out. Maybe the context gets a little skewed the further you get down in the chat. So then I'll just take whatever it had there and then my problem, and then start a new chat.}''
\end{itemize}

These six strategies reveal sophisticated prompt engineering practices among early adopters, demonstrating that effective LLM use requires deliberate experimentation, iterative refinement, and careful context management. Notably, participants did not converge on a single "best practice" but rather developed flexible, context-dependent approaches that balanced specificity with openness, security with informativeness, and persistence with knowing when to start fresh.

\paragraph{\textbf{Challenges}}
\begin{itemize}
\item \textbf{Struggles with Integration and Adjustments:} Five participants observed that LLMs faced challenges when integrating context and making adjustments or changes. P1 shared an incident in which ChatGPT struggled to change minor components in its output: ``\textit{It wouldn't change just that piece of the code. It would change all of the code.}'' Participants also expressed concerns about LLMs' abilities to retain the original context over multiple prompts or interactions.
\item \textbf{Limitations of Context Window:} Five participants highlighted the limitations of LLMs' context windows. They noted instances where LLMs ran out of space to generate code or lost context when prompted multiple times.
\item \textbf{Necessity of Follow-up Queries and Challenges of Providing Context:} Two participants shared that they often relied on follow-ups because LLMs did not always get things right the first time. One participant noted the challenges of providing sufficient context, especially when unsure of what specific examples to provide.
\end{itemize}
\paragraph{(2) Evaluating Context}
\begin{itemize}
\item \textbf{Verifying through Reading Code:} Ten participants shared that they rely on reading their code as a primary method of review. P9 noted that reading could be a sufficient check for a simple piece of code: ``\textit{For me, as a professional developer, it is still my job to validate [that] this piece of code is going to work, and it's going to integrate well into my existing codebase. So to do that, if it's something that's simple enough, I can just mentally do a quick check --- me reading it line by line.}''
\item \textbf{Verifying through Output:} Similar to testing, ten participants shared that they engage in validating their code by checking the output after reading through the code. As P8 shared, ``\textit{If that output would be contradicted with what I expect, I discard it.}''
\item \textbf{Verifying through Manual Testing:} Seven participants also engaged in manual testing practices, such as verifying code templates and solutions, and using developer tools or REPL tools to test code functionality before integration. Checking console logs, verifying expected function return values, and monitoring IDE warnings or errors were common practices among participants.
\item \textbf{Verifying through External Tests and Sources:} Four participants employed external testing frameworks and tools to evaluate the generated code. Two participants sought validation from external sources such as the Internet, looking up documentation and searching platforms like Stack Overflow.
\item \textbf{Other Verification Methods:} Two participants asked LLMs to explain their own generated code as part of their evaluation process. Two developers working in front-end development noted that they visually assessed the quality of their code. One participant shared how he assessed readability using SOLID principles, clean code concepts, and official documentation. One participant noted ChatGPT's self-correction capability in recognizing and rectifying errors without explicit user guidance.
\end{itemize}
\paragraph{\textbf{Challenges}}
\begin{itemize}
\item \textbf{Increased Skepticism:} Five participants became more skeptical and vigilant about evaluating code quality due to the existence of generated code. P1 shared: ``\textit{[I]n a sense, yes, it has changed testing. I'm a bit more skeptical.}''
\item \textbf{Necessity of Evaluation:} Seven participants emphasized the importance of thoroughly checking the generated code for accuracy and relevance. P6 shared, ``\textit{Be wary of what you get [...] Don't give up on checking the responses.}'' Two participants shared that they paid more attention to reviewing generated code compared to human-written code.
\item \textbf{Lack of Self-Evaluation and Contextual Clarity:} One participant mentioned ChatGPT's lack of self-evaluation as a limitation. Another emphasized the importance of context in understanding code, suggesting that humans have an advantage in being able to provide contextual information about how and why code was created.
\end{itemize}
\paragraph{(3) Integration}
Here, we examine integration across two dimensions. First, we investigate the practices employed by developers to incorporate the generated code into their work. Second, we assess the degree to which developers actually utilized the generated code.
\begin{itemize}
\item \textbf{Modifying Before Integration:} Fifteen participants reported that they modified the generated code before integrating it into their projects. P13 shared her process: ``\textit{Once I got that overall approach, I refined it in a few places, and then I sent it back to ChatGPT [...] 'Kind of give me boilerplate code for that.' So I think it did a really good job at giving me that boilerplate code. And then I just had to do a lot of refinements within it to overcome certain errors.}''
\item \textbf{Copy-Pasting Code:} Twelve participants reported simply copy-pasting the generated code into their development environments. P7 stated, ``\textit{Maybe there [aren't] too [many] steps. Just asking the question, and if the answer looks good to me, I would just copy-paste.}''
\item \textbf{Rewriting Code:} Three participants mentioned that they preferred to rewrite the generated code manually to gain a deeper understanding and mitigate potential errors.
\item \textbf{Using and Discarding Generated Code:} Six participants rarely discarded code, two discarded about half of all generated code, and eight discarded the majority of their generated code. Those who seldom discarded code pointed out that, with well-crafted prompts, LLMs are capable of producing viable code. Eight participants shared that they threw away a lot (greater than 50
\end{itemize}
\paragraph{\textbf{Challenges}}
\begin{itemize}
\item \textbf{Friction with Copy-Pasting:} One participant expressed frustration with the copy-pasting process due to the need to switch between applications or rely on third-party plugins.
\end{itemize}
\subsubsection{\textbf{Testing and Code Review}}
\begin{itemize}
\item \textbf{Generating Unit Tests:} Seven participants had used LLMs for generating tests, particularly for unit tests due to their relatively basic and formulaic nature. One participant noted that using ChatGPT to generate tests encouraged him to incorporate more testing into his coding practices.
\item \textbf{Simulating Code Reviews:} Two participants, who worked independently as contractors or in small teams, used LLMs for code review due to the lack of colleagues to review their code; an additional participant used ChatGPT to review chunks of his code.
\end{itemize}
\paragraph{\textbf{Challenges}}
\begin{itemize}
\item \textbf{Inapplicable for Code Review:} Eight participants indicated that they did not utilize ChatGPT for code reviews or that it had minimal impact on their code review process. P15 expressed hesitance: ``\textit{I don't use it for code review, mainly because I need to understand what the code is doing myself [...] It's better for me to know each step of what it's trying to do, because I need to know ... how it's going to affect the rest of the system.}''
\item \textbf{Struggles with Complex Tests:} Two participants noted that LLMs struggled with generating larger or more system-wide tests due to security considerations or lack of necessary context.
\end{itemize}
\subsubsection{\textbf{Debugging, Refactoring, and Documentation}}
\begin{itemize}
\item \textbf{Improving Debugging:} Ten participants mentioned having used LLMs to debug their own code. Participants highlighted LLMs' efficacy in significantly reducing debugging time by providing quick solutions or guiding them in the right direction. P5 shared: ``\textit{[ChatGPT] helps to dramatically shorten the whole debugging process. If it doesn't give you the answer [...] it helps to put me in some right directions to where I can do some further research or ask it more questions.}''
\item \textbf{Reducing Syntax-Based Errors:} Two participants emphasized LLMs' utility in resolving syntax-related errors, such as missing punctuation or braces.
\item \textbf{Performing Refactoring:} Two participants shared that they used LLMs for refactoring purposes, aiming to enhance maintainability and comprehensibility while preserving functionality. Additionally, one participant utilized LLMs to condense code, particularly by converting code blocks into more concise versions.
\end{itemize}
\paragraph{\textbf{Challenges}}
\begin{itemize}
\item \textbf{Debugging Difficulties:} One participant shared a negative experience while using ChatGPT to debug code found online, emphasizing that his lack of experience with the code left him unable to properly evaluate ChatGPT's output, leading to frustration when ChatGPT mixed up Python syntax with other languages.
\end{itemize}


\subsection{RQ3: How has the use of LLMs influenced the software products created?}
The key themes that emerged in understanding how LLMs affect the artifacts (i.e., code and software products) are as follows:
\subsubsection{\textbf{Quality and Complexity of Generated Code}}
\begin{itemize}
\item \textbf{Producing Clean, Readable Code:} Six participants shared that they found LLMs' generated code to be clean, readable, and systematic. They likened its readability to that of standard Stack Overflow answers or official documentation for programming languages. Five participants, such as P4, appreciated the ease of understanding the syntax: ``\textit{[ChatGPT is] generally pretty good about not generating difficult to read code or overly complex [code]. And just asking it to improve the code in terms of readability generally works for small snippets. I've had good success with that.}''
\item \textbf{Generating Code with Reasonable Complexity:} Five participants stated that the code actually had fair to good complexity for the tasks that they were working on. Particularly for common tasks or well-known problems, participants generally found the complexity to be sufficient. P11 stated that he found ChatGPT to have good complexity for more common tasks, but not for more novel tasks: ``\textit{Every now and then I'll put in a code challenge, or an interview kind of problem, and [...] I think it does better on those, because other people have posted about those kinds of problems. [...] If I try to give it more unique data structure problems from my actual experience, it doesn't do as well.}''
\item \textbf{Conducting Complexity Analysis:} Four participants utilized LLMs as tools for checking the time and space complexity of code snippets and explaining the benefits of different approaches. This served as a learning opportunity, particularly for participants who were newer to computing.
\end{itemize}
\paragraph{\textbf{Challenges}}
\begin{itemize}
\item \textbf{Struggling with Up-to-Date Information:} Two participants noted that ChatGPT's generated code sometimes lacked up-to-date information, particularly when dealing with rapidly evolving technologies or APIs.
\item \textbf{Over-Engineered, Complex Code:} Five participants raised concerns about LLMs over-engineering solutions, adding unnecessary complexity and inefficiency to simple problems. For instance, as P8 complained, ``\textit{One complaint that I have about ChatGPT's output is over-engineering. [...] In some of cases, I'm feeling that sometimes that I'm asking it, `Okay, write a simple method for, I don't know, multiplying two numbers together.' [...] Sometimes it does the over-engineering for those cases.}'' For more novel or unique challenges, some participants observed limitations in ChatGPT's ability to produce code with optimal complexity.
\end{itemize}
\subsubsection{\textbf{Optimal Use Cases}}
\begin{itemize}
\item \textbf{Excelling at Small Tasks:} Nine participants identified LLMs' strength in handling small tasks, particularly those that involve routine or standard procedures. They found it most effective for tasks that could be decomposed to a low level, such as writing small code snippets or implementing basic functionalities. For ChatGPT, P4 shared that ``\textit{I've kind of isolated its use cases to helping me improve code at the function level, like small snippets of code, and helping me go from zero to something}''; this sentiment is connected to his relationship with frequently discarding code as part of an iterative process.
\end{itemize}
\paragraph{\textbf{Challenges}}
\begin{itemize}
\item \textbf{Better at Text than Code:} Two participants observed that LLMs seemed to perform better at generating textual content than code, possibly due to their training data composition \cite{wu2023brief}.
\end{itemize}
\subsubsection{\textbf{Security}}
\begin{itemize}
\item \textbf{Providing Sufficient Security:} Thirteen participants expressed that they considered LLMs' code to be sufficiently secure for their purposes, primarily because they were not using it in production environments or for critical applications. It was also mentioned that LLMs' code is as secure as any publicly available code. Seven participants shared that this was because of their specific use case, like P2, who noted that ``\textit{I've never had to use it for any security aspects.}'' P14 additionally used a VPN, and P4 and P15 shared that they believed their reading of the generated code was another layer of added protection against security concerns.
\end{itemize}
\paragraph{\textbf{Challenges}}
\begin{itemize}
\item \textbf{Concerns Sending Data to LLMs:} Eight participants shared that they specifically did not provide LLMs with identifying, confidential, or otherwise proprietary information in their prompts. P3 expressed concerns over the lack of \textit{trust and safety controls.}'' A few participants noted concerns about their data being used by companies like OpenAI. P12 shared related concerns: \textit{If those interactions with Bard or ChatGPT get logged and used for training data in the future, it could be that those models start outputting production code or like, our own internal code. And that's something we want to avoid.}'' It is worth noting that, since April 2023, ChatGPT has included an option to prevent a user's queries from being used to train or improve the model \cite{dastin2023incognito}; the feature that both P14 and P15 mentioned that they used in order to reduce their security concerns. At present, Google Gemini additionally allows users some control over how their data is used by Google \cite{gemini_privacy}, and Copilot allows users to opt out of sharing their data \cite{copilot_privacy}.
\item \textbf{Concerns using Data from LLMs:} Eight participants emphasized developers' responsibility to ensure code correctness, especially regarding the possibility of copying and using code without review. Two participants explicitly stated that LLM code should not be copied without review due to security concerns; P11 shared that, ``\textit{I think it's dangerous, right? Like, if out of our work on a team, somebody would just copy-paste ChatGPT [code], you know, I'll probably be annoyed by it.}'' P1 further expressed concerns over malicious actors potentially injecting poisoning code that goes on to then further train ChatGPT, which could result in the tool generating exploitable code in the future.
We note that both this and the above developer-identified challenge are clearly informed by applying knowledge of how LLMs work to determine possible risks of sharing data with LLMs and using LLM outputs in production code.
\end{itemize}


\subsection{RQ4: How may the software industry and education be affected by LLMs?}
The key themes that emerged in understanding how LLMs impact two areas of society (the software industry and CS education) are as follows:
\subsubsection{\textbf{Industry}}
\begin{itemize}
\item \textbf{Comparing LLMs to Existing Entities or Roles:} We found nine interviews in which LLMs were fulfilling roles commonly filled by other people or tools, including as a pair-programmer, assistant or secretary, junior developer, rubber duck, or simply a tool. For instance, P10, whose colleagues had been all laid off, mentioned, \textit{I've been working solo by myself for a few months now. [ChatGPT] is my junior who [is] trying to help me.}" P5 also mentioned, \textit{[ChatGPT] is like the rubber ducky that we would have, except now it produces answers, and it talks to me, and it gives me all of the advice that I would need.}'' These results suggest that developers are in the process of figuring out the future of their careers and how LLMs may ultimately fit in.
\item \textbf{Minimal Impact on the Job Market:} Nine participants expressed that jobs in the software development field would remain largely unaffected. They argued that software development involves more than just code writing, and LLMs lack the capability to fully substitute human developers. As P3 noted, ``\textit{I don't really see it replacing people because I think if you're at the point, at least with the current version of it, that it could do your job, you probably weren't as good of a programmer in the first place.}''
\item \textbf{Widespread Use Among Peers:} Four participants observed widespread use of ChatGPT among their peers, suggesting a growing acceptance and adoption of LLMs within professional environments, particularly within tech companies.
\item \textbf{Lowering Entry Barriers:} Three participants mentioned LLMs could lower the barrier for entry-level positions by providing assistance and resources. P6 shared his hopes on LLMs being able to answer novice programmers' questions: ``\textit{[ChatGPT] has lowered the barrier to entry in terms of coding. So you can really go and ask and stuff, and depending on what your level of expertise is or what your level of questioning is, it will provide you with the level of answers.}''
\end{itemize}
\paragraph{\textbf{Challenges}}
\begin{itemize}
\item \textbf{Changing the General Job Market:} Nine participants viewed LLMs as technologies that would change and re-purpose jobs in general. Although they acknowledged LLMs' capacity to diminish some jobs, they noted their potential to create new job opportunities as well. For example, P10 noted, ``\textit{[ChatGPT] is a very powerful tool, and it's going to kill a lot of wild white-collar jobs for sure, but more jobs will be created. [The issue] is just how fast that those jobs will be created and whether the people that lost their job will get trained.}''
\item \textbf{Absence and Necessity of Guidelines:} Eight participants emphasized the importance of establishing guidelines for LLM use, especially for larger companies or those dealing with security-sensitive information. P9, for example, noted, ``\textit{I think 100\% [that companies should have guidelines], especially when you're dealing with things that are sensitive; for example, financial institution or medical [data]. And I actually do think that instead of them trying to use third-party LLMs from, for example, OpenAI, they should start building their own models and have them housed within their own PVC cloud so that it's secure enough for their standard.}'' Of these eight participants, five noted that their companies lacked specific guidelines for LLM use, though existing rules about not sharing sensitive information applied.
\item \textbf{Lowering the Need for Certain Roles:} Five participants highlighted LLMs' potential to decrease the number of developers required for certain tasks and roles, such as user interface professionals or data scientists.
\item \textbf{Undermining Entry-Level Positions:} LLMs' potential to undermine the demand for entry-level roles emerged three times in our study. P6, for instance, mentioned, ``\textit{I do find the level of code generated is sometimes almost as good as a junior software developer. So I think that really up the bar of hiring for junior software developers.}'' P15 also stated that newer developers should broaden their skill set in order to stay ahead of technological advancements.
\item \textbf{Exploiting LLMs in Job Interviews:} Two participants raised concerns about candidates using LLMs to pass interviews without genuine knowledge or skills, suggesting that formulaic interview techniques might need to change.
\end{itemize}
\subsubsection{\textbf{Education}}
\begin{itemize}
\item \textbf{Encouraging as Learning Tools:} Ten participants shared that CS curricula should integrate LLMs into education rather than banning them outright. P5 shared that, ``\textit{If universities are trying to prepare students for work in the industry, I would say that exposing the students to how to properly use ChatGPT to create impactful works in code and program would certainly be beneficial in the professional corporate setting. But, you still need to have the foundational knowledge of your data structures, algorithms, how to design code databases, SDLC, all of those different core topics. Those are still needed within the curriculum.}'' Participants highlighted LLMs' potential to assist students in problem-solving, generating personalized problem sets, and improving prompt engineering skills.
\item \textbf{Teaching Prompt Engineering:} Four participants emphasized the importance of teaching students prompt engineering. For instance, P13 mentioned, ``\textit{I think students should be introduced to these tools and how they can prompt better. For instance, one of my instructors at university, the first thing that they covered in our first semester, was how to Google. I think this is going to be something along the same lines as how to prompt.}''
\item \textbf{Impracticality of Bans:} Three participants viewed the issue from another perspective: the impracticality of banning it. For instance, P2 stated, ``\textit{You cannot run away from it because everyone has access [...] Students are going to use it, no matter [what].}'' Participants emphasized the potential benefits of LLMs in supplementing learning, particularly for underprivileged students who may lack access to traditional tutoring or resources.
\end{itemize}
\paragraph{\textbf{Challenges}}
\begin{itemize}
\item \textbf{Emphasizing Fundamental Concepts:} Six participants stressed that foundational knowledge, theoretical understanding, and fundamental concepts should be prioritized over the integration of ChatGPT. They believed that topics like software architecture, algorithm design, and problem-solving skills should take precedence. For instance, P1 mentioned, ``\textit{I think [universities] should be doing more of what they should have been doing in the first place, which is not necessarily focusing on specific implementations and specific kind of optimal algorithms, but rather on the bigger, more difficult-to-grasp ideas that have to do with architecting software and that have to do with thinking about what it takes to get from a business idea to an actual product.}''
Concerns over LLMs being just the latest tool or technology that may become obsolete also appeared. For example, P6 mentioned, ``\textit{One thing remains constant [in CS education] that you are studying and you're understanding and you're learning the programming languages and you're learning the technology that will be obsolete by the time you get to the market for a job. [...] I think, in general, the program should be focused on more fundamentals because those largely remain the same.}''
\item \textbf{Needing to Adapt:} Six participants acknowledged the flip side of the impracticality of banning LLM usage, suggesting ways that computing education programs should adapt to the new reality with LLMs. Four participants expressed that the ease of plagiarism with LLMs encourages students to use them for ready-made solutions, introducing a need for change. Two participants suggested that instructors mitigate the impact of cheating concerns by redesigning assignments to ensure they require a genuine understanding of concepts through critical thinking and problem-solving skills. P2 shared, ``\textit{You can design the assignments in such a way that even if [students] use [ChatGPT], they need to understand the concepts.}'' Two more participants mentioned the importance of preparing students for coding without LLM assistance.
\end{itemize}
\section{Discussion}
\subsection{The Sociotechnical Transformation of Software Development with LLMs} 

Our investigation of early adopter experiences reveals that LLM integration in software development represents a fundamental sociotechnical transformation, rather than simply adopting a tool. Across our four research dimensions—people, process, product, and society—we observe a consistent pattern: LLMs provide substantial benefits when developers possess specific competencies, work within supportive organizational contexts, and maintain critical evaluation practices. The technology itself is neither inherently beneficial nor harmful; rather, its impact depends on the sociotechnical system within which it operates. This finding aligns with established sociotechnical systems theory in software engineering, which emphasizes that technology adoption succeeds or fails based on the interplay between technical capabilities, human practices, organizational structures, and broader social contexts \cite{baxter2011socio}.

These dimensions are deeply interconnected. Developers' growing confidence and productivity (RQ1) enabled more sophisticated process adaptations (RQ2), but only when coupled with rigorous verification practices that maintained product quality (RQ3). Notably, we observe a productivity-quality paradox: participants reported significant time savings yet frequently discarded generated code (eight participants discarded more than 50\%, two discarded approximately half). This apparent contradiction resolves when we recognize that LLMs fundamentally shift developer effort from generation to evaluation—a transformation from syntactic production to critical assessment that requires a deep understanding to identify subtle errors, architectural mismatches, and security vulnerabilities. This shift has profound implications for developer expertise, organizational workflow, and educational preparation (RQ4), as the profession grapples with redefining what it means to ``write code'' versus ``orchestrate code generation.''

Our findings illuminate three critical themes that transcend individual research questions. First, effective LLM use requires new competencies that differ fundamentally from traditional programming skills—prompt engineering, critical evaluation, and security-conscious integration (detailed in Section \ref{sec:implementation}). Second, LLM utility is highly phase-dependent across the software development lifecycle, with near-universal adoption during implementation but minimal impact on requirements gathering (examined in Section \ref{sec:requirements}). Third, human judgment, collaboration, and domain expertise remain persistently important despite automation of certain tasks. Six participants explicitly noted LLMs' inability to replace human interactions, twelve emphasized that requirements gathering remains fundamentally human-centered, and eight stressed the critical importance of problem decomposition skills that LLMs cannot provide.

These themes underscore that successful LLM integration requires not just technological sophistication but also organizational support, educational reform, and ongoing professional development. We explore each theme in depth below, positioning our findings within the broader empirical literature and deriving practical implications for multiple stakeholder groups.

\subsubsection{The Productivity-Quality Paradox}

As introduced in Section 4.1, the productivity-quality paradox, i.e., developers reporting substantial time savings while discarding much generated code, challenges simplistic narratives about LLM productivity and requires careful interpretation. This pattern reveals that productivity manifests through multiple mechanisms rather than direct code reuse.

LLMs shift developer effort from code generation to evaluation and integration. Productivity emerges through: (1) \textbf{rapid prototyping and iteration}, where developers quickly explore multiple approaches even if individual attempts are discarded, as P4 explained: \textit{``I've kind of isolated its use cases to helping me improve code at the function level, like small snippets of code, and helping me go from zero to something''}; (2) \textbf{learning through examples}, where generated code serves as educational material—eleven participants highlighted LLMs' ability to explain code and demonstrate best practices; (3) \textbf{overcoming cognitive blocks}, where LLMs help developers get ``unstuck'' by suggesting approaches or providing boilerplate; and (4) \textbf{reducing search friction}, where LLMs aggregate information from multiple sources, as P14 noted: \textit{``I don't need to click on each of the links from Google to find the solution [...] [ChatGPT] gives me a summarized version from possibly three, four different sources.''}

This reframing reveals LLMs' role as \textit{thinking partners} or \textit{ideation tools} rather than \textit{code factories}. The value lies in cognitive support, i.e., helping developers think through problems, explore solution spaces, and learn approaches, rather than in generating production-ready code. Fifteen participants modified generated code before integration, and twelve copy-pasted as a starting point for adaptation rather than final implementation. As P12 articulated: \textit{``You can take [Bard's response] as an idea, like a first draft, and then iterate on it; that's the experience I have working with it from the coding side.''}


However, this productivity shift introduces new risks. The ease of generating plausible-looking code may reduce cognitive engagement necessary for deep learning and quality assurance. Five participants reported becoming more skeptical in their code review practices specifically because of LLM-generated code, with P1 noting: \textit{``[I]n a sense, yes, it has changed testing. I'm a bit more skeptical.''} This heightened skepticism is appropriate. The productivity-quality paradox thus reveals both opportunity, i.e., LLMs can accelerate development when used strategically and responsibility, i.e., developers must cultivate stronger evaluation skills to ensure productivity gains do not compromise code quality, security, or learning.

\subsection{Positioning Our Findings: Early Adoption Patterns and Subsequent Validation}

Our findings as early-adopter data collected in mid-2023 provide a valuable lens into emergent patterns that subsequent large-scale research has also reported. The convergence between our qualitative insights and subsequent quantitative studies demonstrates both the reliability of our findings and the value of early adopter perspectives in understanding technological adoption trajectories.

Several key findings have been corroborated by large-scale industry research conducted between 2024 and 2025. Our participants' reports of substantial productivity benefits align with GitHub and Accenture's 2024 study finding that 96\% of developers report faster task completion with AI tools, with an average productivity increase of 55\% \cite{github_accenture_2024}, further corroborated by a systematic literature review carried out by Hou et al \cite{hou_large_2024}. The security concerns we documented align with Stack Overflow's 2024 developer survey, which identifies security as the top concern about AI tools, with 45\% of developers worried about data privacy implications \cite{stack_overflow_survey_2024}, further corroborated by Islam. et al in their 2024 work on LLM induced vulnerabilities \cite{islam_vuln_2024}. Our finding that LLMs facilitate learning is echoed in recent educational research, which demonstrates LLMs' effectiveness as personalized learning tools while raising concerns about potential learning impediments \cite{sajadi2025llms}. The prompt engineering strategies we documented have since been formalized in training materials and research on effective prompting patterns \cite{cui2025effects}.

This convergence validates our early-adopter methodology: patterns observed among pioneers who adopted LLMs between November 2022 and April 2023 proved predictive of broader adoption trends visible in studies with thousands of participants conducted 12 to 18 months later. Early adopters serve as ``scouts'' for technological change, developing practices and encountering challenges before they become widespread. Innovation diffusion theory identifies them as typically more educated, higher in social status, and more willing to tolerate uncertainty \cite{rogers2014diffusion}. Our participants exemplify these characteristics, developing sophisticated practices through intensive experimentation that later adopters can now learn through structured training.

However, convergence alone understates our contribution. Our qualitative approach and comprehensive SDLC focus reveal insights less visible in survey-based or tool-specific studies, making three distinctive contributions to the literature. First, our phase-dependent adoption analysis, showing minimal requirements-gathering impact (12/16) despite near-universal implementation adoption (16/16), provides granular insight into \textit{where} and \textit{why} LLMs prove useful across the development lifecycle. Most existing studies focus narrowly on coding tasks \cite{Tang2024MLBench} or specific tools \cite{fatima2024flakyfix}, while our comprehensive SDLC mapping reveals that LLM utility is highly context-dependent, challenging oversimplified narratives about LLMs ``transforming software development.''

Second, our taxonomy of six prompt engineering strategies and five verification methods, derived from observed practice rather than prescribed techniques, provides empirical grounding for understanding how developers actually use LLMs. Our early documentation captures the emergent, experimental nature of these practices before they became formalized, revealing not just \textit{what} works but \textit{how} developers discovered effective approaches through trial and error.

Third, our finding that high code discard rates (8/16 discard >50\%) coexist with productivity gains challenges simplistic ``productivity improvement'' metrics prevalent in industry reports. This productivity-quality paradox reveals that LLMs' value lies in cognitive support (ideation, exploration, learning) rather than direct code reuse. This nuance, which requires a qualitative investigation to uncover, fundamentally reframes how we should measure and understand LLM ``productivity.'' Survey-based studies asking ``Do you find LLMs productive?'' capture perceived value but miss the complex reality of substantial code rejection alongside productivity benefits.

Our study's distinctive value emerges from the intersection of three methodological choices: (1) \textbf{timing}, capturing early adoption before practices became standardized, (2) \textbf{method}, qualitative depth revealing nuances invisible in surveys, and (3) \textbf{scope}, comprehensive SDLC coverage rather than narrow task focus. By documenting early-adopter experiences during the formation stage of LLM integration practices, our study captures not just outcomes but processes, the reasoning, experimentation, and sense-making through which effective practices emerged. This developmental perspective is valuable for supporting later adopters, who benefit from knowing not only \textit{what} to do but also \textit{why} certain approaches work and \textit{how} to adapt them to specific contexts.

\subsubsection{For Individual Developers: New Competencies and Practices}

Effective LLM integration requires developers to cultivate competencies that differ from traditional programming skills. Our findings reveal three critical capability clusters: (1) prompt engineering and problem formulation, (2) critical evaluation and verification, and (3) security-conscious integration. These competencies complement rather than replace traditional skills; developers still need strong fundamentals in data structures, algorithms, and software architecture to evaluate outputs, identify errors, and make sound design decisions. As P5 emphasized, LLMs require \textit{``a solid foundation to pose the right queries.''}

Our findings provide substantial clarification to a distinct, role-based divergence in both LLM application and perceived risk. We found that junior developers, in particular, derived significant educational value; all junior participants (4/4) felt that access to an LLM helped foster their learning; two framed the LLM as a ``pair-programmer" (P5, P9) that accelerated their onboarding and understanding of unfamiliar code. Conversely, senior developers primarily leveraged LLMs to streamline software development workflows by offloading mundane tasks, thereby preserving their cognitive resources for complex architectural challenges. However, our study also complicates this narrative by identifying a tension in perceived risk. While junior developers focused on the immediate learning benefits, it was predominantly senior developers (5/8 participants in this theme) who articulated fears of stagnation of skill development, expressing explicit concern that over-reliance could atrophy the foundational problem-solving skills of their less-experienced colleagues. Our work thus refines the "role-dependent" hypothesis by specifying how use cases diverge and revealing that concerns over skill erosion are themselves stratified by seniority.

\paragraph{Prompt Engineering and Problem Formulation}
Prompt engineering emerged as a sophisticated practice encompassing six distinct strategies. Fifteen participants varied the specificity of prompts based on desired outputs, recognizing that broad prompts encouraged creative solutions, while specific prompts yielded targeted answers. As P15 explained: \textit{``[I]t started to give me actually better results, because it knew things that I didn't know [...] When I have a more broad [prompt], then it's able to kind of formulate its own deduction.''} Conversely, P2 advocated specificity for software tasks: \textit{``I think it should be very specific [...] for software development, I think you need to be very specific.''} The strategic lesson is that effective prompters adapt their specificity to context.

Beyond specificity, participants employed iterative refinement (fifteen used follow-up queries), security-conscious generalization (thirteen avoided sharing confidential information), contextual examples (nine practiced few-shot prompting), conversation management (nine started new threads when context degraded), and deliberate practice (fourteen acknowledged prompt engineering as essential). As P16 noted: \textit{``The follow-up questions actually might be more important than [the initial question].''}

These strategies require metacognitive awareness; understanding not just the problem but also how to formulate it for an LLM's particular strengths and limitations. This represents a shift from directly solving problems to \textit{orchestrating problem-solving}, a form of meta-programming that complements traditional coding skills.

Closely related is problem decomposition. Eight participants noted that LLMs work best with small, well-defined tasks, necessitating more deliberate decomposition. Interestingly, some viewed this as a ``forcing function'' encouraging better practices. P4 shared: \textit{``[ChatGPT] might force me to break [problems] down [...] If I can't explain it to ChatGPT, then it might be an indicator that I don't yet understand it.''} This suggests that LLM constraints may inadvertently reinforce fundamental SE principles.

Developers can build prompt engineering competencies through formal resources such as OpenAI's comprehensive guide \cite{openai_prompt_nodate} or DeepLearning.AI's course \cite{deeplearningprompt}, though experiential learning, i.e., deliberately practicing on non-critical tasks and reflecting on what works, is equally valuable for developing the metacognitive awareness that characterizes expert LLM use.

\paragraph{Critical Evaluation and Verification}
Our participants employed five verification methods to ensure code quality: reading generated code (ten participants), checking outputs against expectations (ten), manual testing using developer tools or REPL environments (seven), external testing frameworks (four), and consulting external sources, such as documentation (two). These practices reveal a multi-layered verification approach where developers combine automated and manual techniques to build confidence in generated code.

Five participants reported becoming more skeptical and vigilant in their code review practices, particularly due to LLM-generated code. As P6 advised: \textit{``Be wary of what you get [...] Don't give up on checking the responses.''} This heightened critical stance is appropriate: LLM-generated code can appear syntactically correct while harboring subtle logical errors, using outdated APIs, or introducing security vulnerabilities. Two participants explicitly stated that LLM code should not be copied without review, with P11 noting: \textit{``I think it's dangerous, right? Like, if out of our work on a team, somebody would just copy-paste ChatGPT [code], you know, I'll probably be annoyed by it.''}

The verification burden explains our productivity-quality paradox: developers save time during generation but must invest substantial effort in evaluation. Fifteen participants modified code before integration, making substantive adaptations to fit their specific contexts, correcting errors, and aligning with project standards. This pattern reveals that LLM's ``productivity'' manifests through rapid iteration and exploration rather than direct code reuse, requiring developers to possess the expertise necessary to critically assess and adapt generated outputs.

\paragraph{Security-Conscious Integration}
Security-conscious integration emerged as a critical practice, with participants demonstrating bidirectional awareness: protecting proprietary data when prompting and scrutinizing generated code for vulnerabilities. Thirteen participants generalized prompts or used case-specific information rather than project-specific details. As P2 explained: \textit{``Because I cannot put all the information into ChatGPT, because it's very client-specific information, it's confidential information [...] I just give it use cases that are similar.''}

This data protection practice reflects appropriate caution: code and prompts inputted into LLM tools may be collected and used by companies to improve their services. Developers must ensure contract compliance and address licensing issues to mitigate risks. Some participants employed additional protections; P14 used a VPN, while P4 and P15 relied on careful code review as a security layer, but these individual measures cannot substitute for clear organizational policies. 

On the output side, eight participants expressed concerns about using LLM-generated code without thorough review. Since LLMs often do not provide the source of generated code \cite{dou2024s}, the origins and security remain unclear, particularly concerning for production environments. P1 expressed concerns about malicious actors potentially injecting poisoning code that could propagate through training, i.e., a supply chain security concern extending beyond individual developer practices.

Thirteen participants considered LLMs' code sufficiently secure for their purposes, but this confidence was context-dependent: participants primarily used LLMs for learning, prototyping, or non-production development. As P2 noted: ``I've never had to use it for any security aspects.'' This underscores that security practices must be calibrated to use cases, with more stringent standards for production code.

Key mitigation strategies include: (1) \textbf{sanitizing generated code} by carefully reviewing and cleaning LLM outputs to prevent vulnerabilities; (2) \textbf{maintaining data integrity} by ensuring confidentiality of proprietary code and sensitive data; and (3) \textbf{upholding security standards} by consistently applying established security practices to avoid compromising systems or violating regulations \cite{islam_vuln_2024}.

\paragraph{Cultivating Critical LLM Literacy}
Underlying these competencies is critical LLM literacy, discerning when, how, and to what extent to trust LLM outputs. Developers must recognize inherent limitations: hallucinations (6/16), knowledge gaps (5/16), struggles with novel problems (4/16), and inability to provide citations (11/16 desired references). Our findings reveal that LLMs excel at boilerplate generation (10/16), code explanation (11/16), and debugging guidance (10/16), but struggle with requirements gathering (12/16 found no impact), novel problems, and complex reasoning.

Structured courses or workshops can accelerate this learning \cite{reichert2024empowering}, though experiential practice on non-critical tasks remains valuable for developing the metacognitive awareness that characterizes expert use.

\paragraph{Special Considerations for Solo Developers}
Solo developers face unique challenges. Two participants (P9, P10) used LLMs as surrogate team members, or in other words, virtual colleagues for requirements refinement, research, and code review. As P10 mentioned: \textit{``I've been working solo by myself for a few months now. [ChatGPT] is my junior who [is] trying to help me.'' }While not substituting for human collaboration, LLMs can enhance independent developers' self-sufficiency, though vigilance against over-reliance remains essential.

\subsubsection{For Organizations: Governance, Guidelines, and Cultural Adaptation}

While individual developer competency is necessary, organizational support is equally critical for effective and responsible LLM integration. Our findings empirically confirm that the legal and compliance ambiguities identified in the literature are not theoretical but represent immediate, practical challenges for early adopters. The themes of security and privacy concerns were pervasive, with 13 of 16 participants explicitly expressing concerns over leaking proprietary data. This finding directly operationalizes the literature's concerns regarding IP attribution and data privacy. Furthermore, our participants were not passive in this regard; they demonstrated active, bottom-up risk mitigation by employing "security-conscious generalization" in their prompts and actively using tool features to opt out of data training. This developer-led effort was coupled with an explicit organizational void, as 8 of 16 participants highlighted the absence and necessity of guidelines. This result substantiates the "unresolved compliance issues" cited in prior work, revealing that professionals are currently navigating a high-risk legal landscape largely without the formal governance frameworks they deem necessary.

\paragraph{Establishing Clear Guidelines and Policies}
Organizations should establish explicit LLM usage policies addressing four critical dimensions. First, \textbf{data sharing policies} must clarify what information can be shared with external LLM providers. Many participants appropriately generalized prompts to avoid sharing confidential information (13/16), but formalized policies provide clarity and reduce the cognitive burden of case-by-case risk assessment. As P9 emphasized, organizations handling sensitive data require stricter standards: \textit{``I think 100\% [that companies should have guidelines], especially when you're dealing with things that are sensitive; for example, financial institution or medical [data].''}

Second, \textbf{code review requirements} must specify scrutiny levels for LLM-generated code before production deployment. Fifteen participants modified code before integration, and eight expressed concerns about using unreviewed LLM outputs. Organizations might establish different review tiers: a lightweight review for non-critical components, a rigorous review for security-sensitive code, and a potential prohibition for safety-critical systems.

Third, \textbf{approved and prohibited use cases} should be explicitly defined. Participants noted that LLMs work well for certain tasks (boilerplate generation, debugging guidance) but poorly for others (novel problems, complex architecture, requirements gathering). Organizations can leverage these insights to guide appropriate use, restricting LLM use for security-critical components while encouraging it for routine tasks.

Fourth, \textbf{documentation and attribution requirements} should clarify expectations when LLMs contribute to codebases. Organizations might require developers to note when significant code sections originate from LLMs, supporting future maintenance and enabling tracking of LLM-related issues. Five participants noted that their companies lacked specific guidelines, though existing rules about not sharing sensitive information provided some guardrails. Organizations that establish clear, well-reasoned policies position themselves to capture LLM benefits while mitigating risks.

\paragraph{Providing Technical Infrastructure}
Beyond policy, organizations can provide technical infrastructure that supports the effective use of LLMs while maintaining security and control. Participants noted friction when switching between LLM interfaces and development environments. Seamless integration through tools like GitHub Copilot, Amazon Q, and IBM Watsonx Code Assistant reduces workflow disruption and encourages appropriate use by lowering barriers to access.

Organizations with specialized needs might invest in tailored LLM solutions that address concerns about proprietary information while providing contextually-relevant assistance. Retrieval-augmented generation (RAG) \cite{lewis2020retrieval} enables LLMs to dynamically incorporate project documentation, code repositories, and communication threads into their responses, providing more accurate and context-aware suggestions than generic models. As P13 suggested, tools trained on project documentation could particularly aid developers working with legacy or brownfield codebases, while locally hosted models that never transmit data externally can address the data sovereignty concerns eight participants expressed.

These infrastructure investments serve dual purposes: they support developers in using LLMs effectively while providing organizations with visibility and control over LLM usage patterns. Centralized tools enable monitoring of adoption rates, identification of effective practices worth sharing, and detection of potential misuse or security concerns.

\paragraph{Fostering Cultural and Workflow Adaptation}
LLM integration necessitates cultural adaptation beyond policy and infrastructure. Nine participants compared LLMs to team members—pair programmers, junior developers, or assistants—suggesting that teams must reconceptualize workflow and collaboration patterns. As P15 noted, LLMs enable focus on \textit{``higher-level stuff that would take more understanding of the infrastructure,''} suggesting potential elevation of human work toward more strategic, architectural, and collaborative activities.

This shift affects multiple organizational dimensions. \textbf{Hiring criteria} may evolve, with some participants noting that LLMs might raise the bar for junior positions as basic syntax knowledge becomes less differentiating. Organizations should emphasize problem-solving capabilities, communication skills, and critical thinking rather than focusing narrowly on coding speed. However, nine participants believed the overall job market would remain stable, arguing that software development involves dimensions beyond code writing that LLMs cannot address.

\textbf{Onboarding practices} must evolve to include LLM competency training alongside traditional skills. New developers need guidance on prompt engineering, verification methods, and appropriate use cases, i.e., the competencies our participants developed through months of experimentation. Organizations can accelerate this learning through structured training, mentorship from experienced LLM users, and documented best practices.

\textbf{Code review norms} require recalibration. Reviewers must be vigilant about LLM-generated code, which may appear clean while harboring subtle logical errors, using outdated patterns, or introducing security vulnerabilities. As P11 cautioned: \textit{``I think it's dangerous, right? Like, if out of our work on a team, somebody would just copy-paste ChatGPT [code], you know, I'll probably be annoyed by it.''} Organizations should cultivate a culture where admitting to LLM use is acceptable so that reviewers can apply appropriate scrutiny.

Finally, organizations should foster \textbf{communities of practice} where developers share LLM strategies, lessons learned, and effective approaches. Fourteen participants acknowledged that prompt engineering was essential to their success, yet many developed these skills independently through trial and error. Systematic knowledge-sharing, through documentation, training sessions, or internal forums, can accelerate collective learning and establish organizational standards for quality and security \cite{ferino_novice_2025}.

\subsubsection{For Educators: Curriculum and Pedagogy}

Computing education faces a fundamental dilemma: LLMs enable students to generate code without understanding it, potentially undermining learning objectives, yet they are ubiquitous tools that students will encounter professionally and are already using regardless of institutional policies. Our participants' perspectives offer guidance: ten advocated integrating LLMs into education rather than banning them, while six emphasized that fundamental concepts must remain central to curricula. This balance, i.e., embracing LLMs as learning tools while ensuring deep conceptual understanding, defines the educational challenge. 

Our findings directly address the precise literature gap regarding practitioners' perceived effects of LLMs on education. While prior work has focused on the student-centric risks of over-reliance, our study captures the missing practitioner perspective, confirming that early adopters see an urgent need for reshaping computing education (a theme endorsed by 9 of 16 participants). Our participants' views operationalize the abstract concerns from the literature: they advocate for curricula to de-emphasize the "commodity skills" (e.g., syntax, boilerplate) that LLMs now automate, and instead intensify focus on "fundamental concepts" like problem-solving, algorithms, and critical thinking. Furthermore, our results extend this discussion beyond just what to de-emphasize by identifying the specific new skills practitioners demand.

\paragraph{The Impracticality and Inequity of Prohibition}
Three participants explicitly noted the impracticality of prohibiting LLM use. As P2 stated: \textit{``You cannot run away from it because everyone has access [...] Students are going to use it, no matter [what].''} Prohibition creates multiple problems. First, it is \textbf{unenforceable}: students can easily access LLMs on personal devices, making bans ineffective without intrusive monitoring. Second, it \textbf{disadvantages compliant students}: those who follow bans are left unprepared for professional practice where LLM use is standard, while peers who violate bans gain unfair advantages and valuable experience. Third, it \textbf{misses educational opportunities}: LLMs can democratize access to educational support, benefiting students who lack resources for tutoring, as P6 noted: ``[ChatGPT] has lowered the barrier to entry in terms of coding.''

Rather than fighting inevitable technology adoption, educators should channel it productively. The challenge is how to design curricula that leverage LLM benefits while ensuring genuine learning, acknowledging that some traditional exercises lose pedagogical value when LLMs can complete them instantly.

\paragraph{Curriculum Content: Fundamentals and LLM Literacy}
Effective curricula must balance two objectives: preserving fundamental knowledge and cultivating LLM literacy. Six participants emphasized that foundational concepts, such as data structures, algorithms, software architecture, and problem-solving, remain critical. As P1 stressed, universities \textit{``should be doing more of what they should have been doing in the first place, which is not necessarily focusing on specific implementations [...] but rather on the bigger, more difficult-to-grasp ideas that have to do with architecting software.''}

These fundamentals are not threatened by LLMs; they are \textit{essential for} using LLMs effectively. Our findings reveal that developers need a deep understanding to evaluate LLM outputs critically, identify subtle errors, make architectural decisions, and formulate problems effectively. Students who lack foundational knowledge cannot assess whether LLM-generated code is correct, efficient, or appropriate. As P5 emphasized, LLMs require \textit{``foundational knowledge of your data structures, algorithms, how to design code databases, SDLC, all of those different core topics.''} The strongest argument for teaching fundamentals is that LLMs make them \textit{more important}—students need robust conceptual understanding to serve as critical evaluators of generated code.

Simultaneously, curricula should explicitly teach LLM literacy, encompassing an understanding of how LLMs work, their capabilities and limitations across various tasks, prompt engineering techniques, and critical evaluation practices. This is analogous to teaching \textit{``how to Google effectively,''} as P13 noted—a meta-skill about effectively using available tools. Four participants emphasized the importance of teaching prompt engineering, recognizing it as a valuable professional competency.

LLM literacy curricula might include modules on: (1) how LLMs work—basic understanding of training processes and probabilistic nature to manage expectations; (2) capabilities across the SDLC—understanding where LLMs excel (implementation, debugging) and struggle (requirements, novel problems); (3) prompt engineering practice—developing the six strategies our participants employed; (4) verification techniques—teaching the five methods our participants used; and (5) security and ethical considerations—understanding data privacy, code provenance, and responsible use. Importantly, LLM literacy should be integrated throughout the curriculum, rather than taught as a separate course, to demonstrate its appropriate use in various contexts.

\paragraph{Pedagogical Approaches: Assignment Redesign and Authentic Assessment}
Pedagogy must evolve alongside curriculum. Two participants suggested redesigning assignments to require genuine understanding even when LLMs are available. As P2 stated: \textit{``You can design the assignments in such a way that even if [students] use [ChatGPT], they need to understand the concepts.''}

Several pedagogical strategies can achieve this goal. First, \textbf{emphasize explanation and justification} over code submission. Assignments might ask students to implement solutions, then explain design decisions, analyze trade-offs, and justify choices. LLMs can generate implementations, but explaining \textit{why} requires understanding. Second, \textbf{use process-oriented assessment}. Require students to document their development process, including initial problem understanding, challenges encountered, how they utilized LLMs, and what they learned. This shifts evaluation from product to process, making LLM use transparent and assessable.

Third, \textbf{employ oral examinations or live coding} to assess understanding. If students cannot explain code when questioned or cannot solve similar problems without assistance, this reveals insufficient learning. Fourth, \textbf{design integration-focused assignments} that require students to combine LLM outputs with original work, demonstrating synthesis skills. Fifth, \textbf{create assignments requiring novel problem-solving} where LLMs struggle. Our findings reveal that LLMs struggle with novel ideas and complex logical reasoning (4/16 noted struggles), so assignments that require creative problem-solving emphasize higher-order thinking.

Finally, \textbf{teach students to use LLMs as learning tools} rather than solution generators. Educators can model appropriate use: asking LLMs to explain concepts, generate practice problems, or identify errors in student code. When students view LLMs as tutors rather than assignment completers, they engage more deeply with the material. Ten participants advocated encouraging LLM use as learning tools, recognizing their potential to support rather than subvert education when used intentionally.

\paragraph{Addressing the Learning Impediment Concern}
Four participants raised concerns about LLMs potentially impeding learning, particularly for those who may lack expertise to evaluate generated code. This risk is real: students might use LLMs to complete assignments without developing genuine understanding. However, this risk exists with any educational technology—the solution is not prohibition but deliberate pedagogical design. This requires: (1) scaffolded introduction where students first learn fundamentals, then gradually incorporate LLM use; (2) explicit instruction on when and how to use LLMs effectively for learning; (3) assessment redesign that rewards understanding over code production; and (4) metacognitive development where students recognize when LLM use is helping versus hindering their growth.

As P5 articulated: \textit{``If universities are trying to prepare students for work in the industry, I would say that exposing the students to how to properly use ChatGPT to create impactful works in code and program would certainly be beneficial in the professional corporate setting. But, you still need to have the foundational knowledge.''} This balanced perspective—embracing LLMs as professional tools while maintaining rigorous foundational education—should guide curricular reform.

\subsection{The Evolving LLM Landscape: Progress and Persistent Challenges}

Since our interviews were conducted during the Spring and Summer of 2023, the landscape of LLMs has notably evolved. Understanding these changes is crucial for evaluating the continued relevance of our findings and identifying which limitations have been addressed versus those that persist.

\subsubsection{Persistent Limitations and Implications for Longevity}

Despite impressive advances in commercially available large language models from major AI companies such as OpenAI, Google, and Anthropic, which include massively expanded context window sizes, reasoning capabilities, knowledge currency, and performance, several categories of limitations still persist, as empirically validated by recent 2025 studies. \textbf{Technical limitations} include: (1) struggles with truly novel or highly specialized problems where training data is sparse; (2) maintaining perfect consistency across very long, complex interactions; (3) providing citations or transparency about information sources—eleven participants desired references, and this persists as LLMs rely primarily on parametric knowledge; and (4) avoiding subtle semantic errors that appear syntactically correct, requiring continued human vigilance \cite{ferino_novice_2025, Zhang2025SWEBenchLive, hou_large_2024, Tang2024MLBench}.

Moreover, while flagship models have addressed many limitations, smaller or locally deployable models that many organizations favor for privacy or cost-efficiency continue to exhibit more pronounced constraints. Organizations that cannot utilize cloud-based LLMs due to data sovereignty concerns face distinct trade-offs, making our findings on verification practices and security-conscious integration particularly relevant. 

For organizations prioritizing the scale and efficiency that large language models can provide, the literature reveals a major constraint in LLMs' ability to handle large, repository-level contexts, systematically quantified by a new generation of benchmarks \cite{Zhang2025SWEBenchLive, Tang2024MLBench}. The findings from ML-BENCH are particularly revealing: they demonstrate that even reasoning models have a low absolute success rate on these repository-scale tasks, which are the norm in modern corporate contexts. In direct response to this "context-gap," a new paradigm of "Agentic Software Engineering" is emerging, which moves beyond simple code generation toward achieving complex, goal-oriented software engineering objectives \cite{Hassan2025AgenticSE}.

More fundamentally, many limitations we documented are \textbf{sociotechnical rather than purely technical}, rendering them resistant to model improvements alone. LLMs' potential to impede learning (4/16 expressed concern) stems from \textit{how they are used} rather than their technical capabilities—even perfect code generation could undermine learning if students do not engage in problem-solving. Similarly, LLMs' inability to replace human collaboration, stakeholder communication, and domain expertise (12/16 noted this for requirements gathering) reflects the inherently human nature of these activities. Security and privacy concerns around data sharing (8/16) remain salient as long as LLMs are trained on user inputs or hosted externally. The expertise gap between novice and expert users persists—our participants developed sophisticated strategies through intensive experimentation, while later adopters still face substantial learning curves.

The relevance of our findings extends beyond the specific models used by our participants for two reasons. First, \textbf{the patterns we documented reflect fundamental characteristics of software work and LLM-mediated development} rather than transient model weaknesses. The phase-dependent adoption pattern—high utility in implementation and debugging, with minimal impact on requirements gathering—persists because it reflects the nature of these activities (technical versus collaborative, well-defined versus ambiguous) rather than limitations of the model. The productivity-quality paradox, i.e., time savings coexisting with high code discard rates, reveals how LLMs shift cognitive work rather than eliminate it, a pattern that remains relevant as long as human oversight is necessary.

The competencies we identified—prompt engineering, critical evaluation, security-conscious integration—remain essential across model generations. While specific strategies may evolve, the underlying skill of formulating problems effectively for LLM consumption persists. The verification methods our participants employed remain necessary as long as LLMs can generate plausible but incorrect code.

Second, \textbf{early adopters establish practices that become institutionalized as technologies mature}. The six prompt engineering strategies that our participants developed through experimentation are now featured in formal training materials. Their verification workflows inform organizational policies. Their security concerns shape regulatory discussions and vendor features. By documenting these emergent practices during the formation stage, our study captures not only outcomes but also the reasoning and learning processes that shaped effective approaches—insights that remain valuable for understanding how to utilize LLMs effectively, even as the tools evolve.

Future developers integrating new LLM capabilities or organizations adopting LLM technologies will face similar challenges: learning what works through experimentation, developing verification strategies, balancing productivity with quality, and establishing appropriate governance. Our participants' experiences provide a valuable template for navigating these challenges, making our findings a roadmap for LLM integration rather than merely a snapshot of 2023 tool capabilities. As P6 anticipated, \textit{``The fundamentals remain the same.''} While LLM capabilities improve, the human work of problem formulation, critical evaluation, collaborative communication, and responsible integration persists.
\section{Threats to validity}
As a qualitative study, our research faces several threats to validity that should be considered when interpreting the findings.
\subsection{Internal Validity:}

The study's timeline overlapped with a period of rapid change in LLM tools (e.g., early versions of GPT-3.5 and Google Bard), which may have influenced participants’ experiences in ways we could not fully control. Because interviews were conducted over several months, different participants may have interacted with slightly different model versions or configurations, making it difficult to separate tool evolution from differences in individual experience, tasks, or organizational context. As a result, our findings should be interpreted as characterizing early LLM-assisted development in general, rather than the behavior of any specific model.

Additionally, we did not administer a pretest to systematically assess participants' baseline proficiency with LLM tools. We partially addressed this by restricting our sample to full-time software professionals and post-survey questions about industry experience. However, without a uniform baseline measure of LLM-specific skill, some observed differences in practices or outcomes may reflect pre-existing expertise rather than effects of the tools themselves.

\subsection{External Validity:}

The geographic limitation of participants, with the majority based in the US, poses a significant threat to external validity. This limitation restricts the generalizability of the findings, as developers in other regions or cultural contexts may encounter different experiences and challenges. Factors such as localized programming practices, language preferences, and access to specific LLM features can all influence how these tools perform in non-US settings.

Moreover, the lack of diversity in the participant pool, predominantly composed of White or Asian males with over 3 years of experience, further exacerbates the threat to external validity. The experiences and challenges faced by underrepresented groups, including women, non-binary individuals, developers with disabilities, and those from various racial, ethnic, and experience backgrounds, may differ significantly from those of the current sample. Although our study was announced openly on LinkedIn, we recognize the need to implement additional strategies in future research to ensure a more diverse and representative participant pool.

\subsection{Construct Validity:}

Our constructs, such as ``productivity,'' ``learning,'' and ``impact on process,'' are based on participants' self-reports in interviews rather than direct observation of work practices or objective productivity metrics. As a result, our findings may be influenced by recall bias, social desirability, or participants’ own interpretations of these constructs. We attempted to mitigate this by asking for examples, and triangulating across multiple participants, but our conclusions still reflect perceived rather than directly measured effects. A second threat is the rapid evolution of LLM tools. Our results therefore speak most directly to LLM coding assistants in 2023, and should be interpreted as patterns of LLM-mediated work rather than properties of any specific current model. Finally, only a minority of participants used tools such as Google Bard or GitHub Copilot Chat, so tool-specific observations, especially about less-used systems, should be treated as tentative. Our primary focus is on cross-tool practices and perceptions, rather than comparative evaluation of individual products.

\subsection{Conclusion Validity:}

With 16 participants and approximately 16 hours of interview data, our study was designed for in-depth thematic analysis rather than statistical generalization. This sample size is consistent with qualitative recommendations for achieving thematic saturation in relatively homogeneous populations, but it nonetheless limits our ability to make strong claims about prevalence or effect sizes in the broader developer population. We mitigated threats to conclusion validity through standard qualitative techniques, including independent coding by multiple researchers, iterative refinement of the codebook, and consensus-building discussions. Theme counts (e.g., how many participants mentioned a given practice) should therefore be interpreted as descriptive signals within our sample, not as precise estimates of how common those practices are in industry at large.

\section{Conclusion}
It has been clear since LLMs' inception that they would impact software development in different ways. Our study aimed to examine the effects of LLMs on software developers, their processes, products, and society at large. Through sixteen interviews with early-adopter developers, we explored their self-reported day-to-day activities, perceptions, and experiences with LLMs. In our qualitative analysis of their responses, we found that:
\begin{itemize}
    \item \textbf{RQ1: People:} LLMs provide developers with numerous benefits, including enhanced productivity, improved efficiency, time savings, streamlined searching, access to templates, and accelerated learning. However, developers also face challenges, such as occasional unreliable LLM responses. 
     \item \textbf{RQ2: Processes:} In the SDLC, LLMs showed minimal impacts on gathering requirements, planning, and refactoring. However, they had mostly positive impacts on ideation, test generation, debugging, and documentation. Developers used various strategies for prompt engineering and evaluating LLM-generated code, such as entering vague prompts or conducting mental checks.
      \item \textbf{RQ3: Product:} LLMs generate readable code and are effective for simple tasks, but they exhibit varying quality across different questions and encounter difficulties with complex tasks.
       \item \textbf{RQ4: Society:} There is a need for formal, proactive guidelines 
       for software developers on the usage of LLMs in the workplace, particularly to promote the ethical and safe use of generative artificial intelligence.
       Additionally, there is a predicted shift in entry-level positions due to LLMs, and LLMs are perceived as being likely to alter or repurpose development-related jobs, rather than eliminating them entirely. Finally, developers hold the ultimate responsibility for the code they deploy, regardless of its source or the process used to create it.

\end{itemize}

The sixteen interviewed developers show an advanced understanding of how LLMs work and their data sources. This understanding influences the decisions that developers make about when, how, and why to use LLMs. Their insights can be used to develop best practices for the use of LLMs in computing education and workforce settings. Since our interviews were conducted in 2023, we explicitly compared them with more recent studies, and found that early reports of productivity gains, security and privacy concerns, phase-dependent use across the SDLC, and prompt-engineering practices have largely persisted, but some initial expectations about code quality, test automation, and the impact on junior roles have been tempered. Overall, our participants viewed LLMs as offering more opportunities than challenges: despite concerns over code quality and limitations, our participants deemed LLMs to be sufficiently useful for developers. Consequently, our findings suggest that LLMs can be a valuable asset in a professional developer's toolbox. As one participant expressed, ``\textit{[ChatGPT] is like calculators being invented. We're going to ban them for a while, and we're going to tell [students] no, no, no, don't use it. And then, once they graduate, they will use it every day.}''

\section*{Declarations}

\subsection*{Funding}
This material is based upon work supported by the Air Force Office of Scientific Research under award number FA9550-21-1-0108 and the National Science Foundation (NSF) under award numbers IIS-2313890, CCF-2006977, and IIS-1917885. Any opinions, findings, and conclusions or recommendations expressed in this material are those of the authors and do not necessarily reflect the view of the AFOSR or NSF.

\subsection*{Ethical approval}
This study was reviewed and approved by the North Carolina State University Institutional Review Board (Protocol \#25913).

\subsection*{Informed consent}
All participants provided informed consent prior to participation, including consent for anonymized quotations to be published.

\subsection*{Author contributions}
Benyamin Tabarsi and Heidi Reichert contributed equally to this work. Material preparation and data collection were led by Benyamin Tabarsi and Heidi Reichert. Initial coding and analysis were conducted by Benyamin Tabarsi and Heidi Reichert, with additional support from Ally Limke during data tagging and Sandeep Kuttal during tag grouping. The first draft of the manuscript was written by Benyamin Tabarsi and Heidi Reichert. Sam Gilson contributed substantially to refining and updating multiple sections of the manuscript, particularly the related work and discussion sections. Sandeep Kuttal and Tiffany Barnes served as supervising authors, providing guidance on study design, analysis, and interpretation of results. All authors (Benyamin Tabarsi, Heidi Reichert, Sam Gilson, Ally Limke, Sandeep Kuttal, and Tiffany Barnes) reviewed, commented on, and approved the final manuscript.

\subsection*{Data availability}
The interview data that support the findings of this study contain information that could compromise participant privacy and are therefore subject to Institutional Review Board (IRB) restrictions (NC State – Protocol \#25913). In keeping with these ethical requirements, full audio recordings and verbatim transcripts are not publicly available.

\subsection*{Conflict of interest}
The authors have no relevant financial or non-financial interests to disclose.

\subsection*{Clinical trial number}
Clinical trial number: not applicable.

\bibliography{_12sam_lib,_11benny_lib,_10ref}

\end{document}